\documentclass[fleqn,usenatbib]{mnras}

\usepackage[T1]{fontenc}
\usepackage{ae,aecompl}
\usepackage{newtxtext,newtxmath}

\DeclareRobustCommand{\VAN}[3]{#2}
\let\VANthebibliography\thebibliography
\def\thebibliography{\DeclareRobustCommand{\VAN}[3]{##3}\VANthebibliography}

\usepackage{graphicx}	
\usepackage{amsmath}	
\usepackage{amsfonts}	
\usepackage[usenames,dvipsnames]{xcolor} 
\usepackage[labelfont=bf]{subcaption} 
\usepackage{hyperref}
\usepackage{array,multirow}  
\usepackage{enumerate}

\newcommand{\msun}{\,\mathrm{M}_{\odot}}

\newcommand{\yr}{\,\mathrm{yr}}
\newcommand{\ev}{\,\mathrm{eV}}
\newcommand{\kev}{\,\mathrm{keV}}
\newcommand{\cm}{\,\mathrm{cm}}
\newcommand{\mhz}{\,\mathrm{MHz}}
\newcommand{\ghz}{\,\mathrm{GHz}}
\newcommand{\erg}{\,\mathrm{erg}}
\newcommand{\second}{\,\mathrm{s}}
\newcommand{\hz}{\,\mathrm{Hz}}
\newcommand{\mpc}{\,\mathrm{Mpc}}
\newcommand{\cmpc}{\,\mathrm{cMpc}}

\newcommand{\invhmpc}{\,h\,\mathrm{Mpc}^{-1}}
\newcommand{\twoi}{\mathrm{II}}
\newcommand{\threei}{\mathrm{III}}
\newcommand{\km}{\,\mathrm{km}}
\newcommand{\myr}{\,\mathrm{Myr}}
\newcommand{\magn}{\,\mathrm{mag}}
\newcommand{\angs}{\,\text{\r{A}}}
\newcommand{\kelvin}{\,\text{K}}
\newcommand{\millikelvin}{\,\text{mK}}
\newcommand{\sr}{\,\text{sr}}
\newcommand{\watt}{\,\text{W}}

\newcommand{\dex}{\,\text{dex}}

\newcommand{\rangeto}{\,$--$\,}

\newcommand{\TS}{T_\mathrm{S}}
\newcommand{\TK}{T_\mathrm{K}}
\newcommand{\xhi}{x_\mathrm{HI}}

\newcommand{\simcode}[0]{{\textsc{21cmSPACE}}}
\newcommand{\polychord}[0]{\textsc{PolyChord}}
\newcommand{\globalemu}[0]{\textsc{globalemu}}

\defcitealias{2016MNRAS.460..417S}{SF16}
\defcitealias{2018ApJ...868...92T}{T+18}
\defcitealias{2019MNRAS.488.3143B}{B+19}
\defcitealias{2024ApJ...961...50S}{SL24}
\defcitealias{2024A&A...690A.108L}{L+24}
\defcitealias{2021ApJ...922...29S}{S+21}

\title[Exploiting synergies between JWST and 21-cm]{Exploiting synergies between JWST and cosmic 21-cm observations to uncover star formation in the early Universe}

\author[Dhandha et al.]
{Jiten Dhandha,$^{1,2}$\thanks{E-mail: \href{mailto:jvd29@cam.ac.uk}{jvd29@cam.ac.uk}}
Anastasia Fialkov,$^{1,2}$
Thomas Gessey-Jones,$^{2,3,6}$ 
Harry T. J. Bevins,$^{2,3}$ 
Sandro Tacchella,$^{2,3}$\newauthor
Simon Pochinda,$^{2,3}$
Eloy de Lera Acedo,$^{2,3}$
Saurabh Singh,$^{4}$
and Rennan Barkana$^{5}$
\\
$^{1}$Institute of Astronomy, University of Cambridge, Madingley Road, Cambridge CB30HA, UK\\
$^{2}$Kavli Institute for Cosmology, Madingley Road, Cambridge CB30HA, UK\\
$^{3}$Astrophysics Group, Cavendish Laboratory, J. J. Thomson Avenue, Cambridge CB30HE, UK\\
$^{4}$Raman Research Institute, Bangalore, Karnataka 560080, India \\
$^{5}$School of Physics and Astronomy, Tel-Aviv University, Tel-Aviv 69978, Israel \\ 
$^{6}$PhysicsX, Victoria House, 1 Leonard Circus, London EC2A 4DQ, UK
}

\date{Accepted XXX. Received YYY; in original form ZZZ}

\pubyear{2025}

\begin{document}
\label{firstpage}
\pagerange{\pageref{firstpage}--\pageref{lastpage}}
\maketitle

\begin{abstract}
In the current era of JWST, we continue to uncover a wealth of information about the Universe deep into the Epoch of Reionization. In this work, we use a suite of simulations with \simcode\, to explore the astrophysical properties of early galaxies and their imprint on high-redshift observables. Our analysis incorporates a range of multi-wavelength datasets including UV luminosity functions (UVLFs) from HST and JWST spanning $z=6\rangeto14.5$, the 21-cm global signal and power spectrum limits from SARAS~3 and HERA respectively, as well as present-day diffuse X-ray and radio backgrounds. We constrain a flexible halo-mass and redshift dependent model of star-formation efficiency (SFE), defined as the fraction of gas converted into stars, and find that it is best described by minimal redshift evolution at $z\approx6\rangeto10$, followed by rapid evolution at $z\approx10\rangeto15$. Using Bayesian inference, we derive functional posteriors of the SFE, inferring that halos of mass $M_h=10^{10}\msun$ have efficiencies of $\approx 1 \rangeto 2\%$ at $z\lesssim10$, $\approx8\%$ at $z=12$ and $\approx21\%$ at $z=15$. We also highlight the synergy between UVLFs and global 21-cm signal from SARAS~3 in constraining the minimum virial conditions required for star-formation in halos. In parallel, we find the X-ray and radio efficiencies of early galaxies to be $f_X = 0.8^{+9.7}_{-0.4}$ and $f_r \lesssim 16.9$ respectively, improving upon previous works that exclude UVLF data. Our results underscore the critical role of UVLFs in constraining early galaxy properties, and their synergy with 21-cm and other multi-wavelength observations.
\end{abstract}



\section{Introduction}

The early Universe remains an exciting frontier due to its observational and theoretical prospects. Observations of the Cosmic Microwave Background (CMB) at redshift $z\approx1100$ ($t_\text{age}\sim0.4\myr$ after the Big Bang) have provided us with a snapshot of the Universe in its infancy, while large-scale spectroscopic surveys like Baryon Oscillation Spectroscopic Survey \citep[BOSS;][]{2013AJ....145...10D} and Dark Energy Spectroscopic Instrument survey \citep[DESI;][]{2024AJ....168...58D} are mapping the nearby Universe in unprecedented detail. The epochs intervening these bookends of our cosmic history, however, remain largely unexplored. Recently, the James Webb Space Telescope \citep[JWST;][]{2006SSRv..123..485G} has opened a new window into the distant Universe, with its high sensitivity and resolution, allowing us to probe some of the earliest galaxies at $z\gtrsim10$ ($t_\text{age}\lesssim 500\myr$) deep into the Epoch of Reionization \citep[EoR; e.g.][]{2025arXiv250100984W}. Prior to the JWST, the Hubble Space Telescope (HST) had provided us with a wealth of information on the high-redshift Universe, including the UV luminosity functions (UVLF) of galaxies and the cosmic star-formation history \citep[e.g.][]{2021AJ....162...47B}.

Now, in the era of the JWST, hundreds of candidate galaxies have been observed at very early epochs \citep[e.g.][]{2023ApJ...948L..14C,2023ApJ...952L...7A,2023ApJ...954L..46L,2023ApJS..265....5H,2024ApJ...964...71H,2024ApJ...969L...2F,2025arXiv250100984W,2025arXiv250315594P}, with the highest spectroscopically confirmed galaxy JADES-GS-z14-0 at $z=14.32$ \citep{2024Natur.633..318C}, alongside a host of distant supermassive black holes, bright AGN and quasars \citep{2023ApJ...953L..29L,2024Natur.628...57F,2024ApJ...966..176Y,2024NatAs...8..126B,2024A&A...691A.145M}. Indeed, the abundance of these bright galaxies at such high redshifts given the small survey volumes exceed those expected by simple extrapolations of the UVLF (e.g. Figure~10 of \citealt{2024ApJ...969L...2F}; or Figure~6 of \citealt{2025arXiv250100984W}), or based on theoretical predictions (e.g. Figure~17 of \citealt{2023ApJS..265....5H}; or Figure~9 of \citealt{2025arXiv250100984W}). Several suggestions have been made in the literature to resolve this discrepancy, including a top-heavy initial mass function (IMF) of early stars \citep[e.g.][]{2022ApJ...938L..10I,2023ApJ...954L..48W,2024MNRAS.529.3563T,2025A&A...694A.254H,2025A&A...696A.157M}, an increased star-formation efficiency \citep{2023MNRAS.523.3201D,2024A&A...690A.108L,2025A&A...696A.157M,2025MNRAS.538.3210B,2025arXiv250505442S}, the presence of hidden AGN \citep{2024JCAP...08..025H}, or star-formation variability/stochasticity \citep[e.g.][]{2023MNRAS.521..497M,2023MNRAS.519..843M,2023MNRAS.526.2665S,2023ApJ...955L..35S,2023MNRAS.525.3254S,2024arXiv240504578K,2024ApJ...975..192G}.

Despite its grand successes, the era of the very first generation of stars at $z\sim20\rangeto50$ ($t_\text{age}\sim50\rangeto200\myr$), the so-called metal-free Population~III stars (Pop~III), will likely be too faint for JWST to detect. The cosmological 21-cm signal, on the other hand, is a powerful probe of this epoch. The signal is a unique signature of neutral hydrogen in the intergalactic medium (IGM) that is sensitive to the thermal and ionization history of the Universe \citep{2006PhR...433..181F, 2012RPPh...75h6901P} resulting from an interplay between the radiation sourced by stars, stellar remnants, or exotic sources. In particular, the 21-cm signal is sensitive to the properties of the CMB/radio background, early Lyman-$\alpha$ sources such as Pop~III stars at Cosmic Dawn \citep[$z\lesssim30$, e.g.][]{2015MNRAS.448..654Y,2019ApJ...877L...5S,2020MNRAS.493.1217M,2022MNRAS.516..841G,2025NatAs.tmp..132G}, heating of the IGM through X-ray photons and Lyman-alpha scattering \citep[$z\lesssim20$, e.g.][]{2004ApJ...602....1C,2007MNRAS.376.1680P,2014Natur.506..197F,2014MNRAS.443..678P,2021MNRAS.506.5479R}, and reionization from stellar/quasar ultraviolet emissions \citep[$z\lesssim10$, e.g.][]{1997ApJ...475..429M,2003ApJ...596....1C}

Detection of this signal is, however, challenging due to its faintness compared to the bright Galactic foregrounds and instrumental noise. Current experiments aimed at the detection of the 21-cm signal typically focus on two summary statistics: the sky-averaged global signal such as EDGES \citep{2018Natur.555...67B}, SARAS \citep{2022NatAs...6..607S}, REACH \citep{2022NatAs...6.1332D}, MIST \citep{2024MNRAS.530.4125M}, RHINO \citep{2024arXiv241000076B}, PRIZM \citep{
2019JAI.....850004P}; or the two-point correlation function of spatial variations (power spectrum) such as MWA \citep{2013PASA...30....7T}, LOFAR \citep{2013A&A...556A...2V}, HERA \citep{2017PASP..129d5001D} and others\footnote{For a comprehensive list of 21-cm experiments, see \href{https://github.com/JitenDhandha/21cmExperiments}{github.com/JitenDhandha/21cmExperiments}, a community-driven public resource which the author invites the readers to contribute to. The list contains details of individual telescopes, their status and associated bibliography for each.}. Future experiments will be able to go a step further and make 21-cm tomographic maps \citep[e.g. SKA,][]{1997ApJ...475..429M}, or probe even deeper into the Dark Ages from moon-based missions \citep[see, e.g.][and references therein]{2024RSPTA.38230068F}.

In 2018, the EDGES collaboration reported a deeper than expected absorption signal at $z\approx17$ \citep{2018Natur.555...67B}. This tentative detection spurred a flurry of exciting theoretical works on early Universe physics. To add to this mystery, measurements of the diffuse cosmic radio background from ARCADE2 \citep{2011ApJ...734....5F} and LWA1 \citep{2018ApJ...858L...9D} hint at an excess above the known CMB spectrum and known Galactic/extragalactic sources. Standard astrophysical mechanisms of galactic emissions \citep{2019MNRAS.483.1980M}, either synchrotron emission from Pop~III supernovae \citep{2019MNRAS.483.5329J} or radio-loud accretion of supermassive blackholes \citep[e.g.][]{2018ApJ...868...63E, 2020MNRAS.492.6086E} could provide such a diffuse contribution. They could also be sourced by exotic agents like annihilating dark matter \citep{2018PhLB..785..159F}, super-conducting cosmic strings \citep{2019JCAP...09..009B}, or primordial blackholes \citep{2022MNRAS.510.4992M,2022MNRAS.517.2454A}. The 21-cm signal, by its nature, is sensitive to the radio background; an excess there would naturally boost the amplitude of the signal \citep{2018ApJ...858L..17F, 2019MNRAS.486.1763F}, thus explaining both the EDGES detection and radio measurements. Although promising, the EDGES detection has been disputed as potential instrumental systematics \citep{2018Natur.564E..32H,2019ApJ...880...26S,2019ApJ...874..153B,2020MNRAS.492...22S} and follow up observations from the SARAS~3 experiment failed to detect the signal \citep{2022NatAs...6..607S}.

Recently, \citet{2024MNRAS.531.1113P} conducted the most complete work to date utilizing multi-wavelength synergies alongside the 21-cm signal to constrain properties of the early Universe. Their work uses the SARAS~3 non-detection of the 21-cm global signal \citep{2022NatAs...6..607S}, the latest HERA upper limits on the 21-cm power spectrum \citep{2023ApJ...945..124H}, and collated measurements of the cosmic radio background \citep[CRB;][]{2018ApJ...858L...9D} and cosmic X-ray background \citep[CXB;][]{2006ApJ...645...95H,2016ApJ...831..185H}. In this work, we extend their analysis by exploring a more flexible star-formation model, and include the UVLF measurements from large HST and JWST surveys as an additional observable. In particular, we constrain the \textit{star-formation efficiency of galaxies} (SFE), an important quantity widely used in analytic galaxy formation models \citep[e.g.][]{2016MNRAS.460..417S,2018ApJ...868...92T,2024ApJ...961...50S}, as well as the efficiency of X-ray and radio sources. In Section~\ref{theory}, we discuss the theoretical framework, including the enhanced SFE model introduced in this work, followed by an introduction of the observational data used in Section~\ref{observational_data}. We then describe our methodology for Bayesian inference in Section~\ref{inference}, and present our results in Section~\ref{results}. We conclude in Section~\ref{conclusions}.

\section{Theory of the early Universe}
\label{theory}

In order to simulate the observables from the early Universe, we use the code 21-cm Semi-numerical Predictions Across Cosmic Epochs \citep[\simcode\footnote{For an overview, see \href{https://www.cosmicdawnlab.com/21cmSPACE/}{cosmicdawnlab.com/21cmSPACE}.},][]{2012MNRAS.424.1335F,2012Natur.487...70V}. We briefly describe the simulations in Section~\ref{overview21cmspace}, but otherwise refer the readers to the most recent code development papers for details \citep{2023MNRAS.526.4262G,2025NatAs.tmp..132G}. In this work, we add two new features to the code: an upgraded Pop~II star-formation efficiency model (described in Section~\ref{enhanced_sfe}), and a calculation of UVLFs (described in Section~\ref{uvlf_model}) for constraining astrophysics using HST/JWST data. All simulations are performed assuming \textit{Planck 2013} best-fit $\Lambda$CDM cosmology \citep[]['Planck+WP']{2014A&A...571A..16P}: ${H_0 = 67.04\km\second^{-1}\mpc^{-1}, \Omega_\text{b}=0.0490, \Omega_\text{c}=0.2678, n_s=0.9619}$.

\subsection{Overview of \simcode}
\label{overview21cmspace}

The simulations begin at $z_\text{max}=50$ with cosmological initial conditions generated using \textsc{CAMB} \citep{2011ascl.soft02026L} for large-scale Eulerian matter overdensity $\delta$ and baryon-dark matter relative velocity/streaming velocity $v_\text{bc}$, and \textsc{Recfast} \citep{2011ascl.soft06026S} for gas kinetic temperature $\TK$ and ionized fraction $x_\text{e}$. The initial conditions are set over a cosmological box split into $128^3$ cubic cells, assuming periodic boundary conditions, and a resolution of $3$ comoving Mpc, giving a total volume of $(384\cmpc)^3$. The fields $\delta$ and $v_\text{bc}$ are evolved forward in time using linear perturbation theory \citep[see, e.g.][]{2016PhR...645....1B}, while $T_K$ and $x_\text{e}$ are evaluated at each step as we describe below. Halo formation is modelled analytically in each pixel using a hybrid dark matter (DM) halo mass function $dn(M_h,z|\delta,v_\text{bc})/dM_h$ \citep[HMF;][]{2004ApJ...609..474B,2011MNRAS.418..906T,2012MNRAS.424.1335F} that depends on the large-scale density fluctuations and streaming velocity resolved at $3\cmpc$ scales. This large-scale overdensity is capped at $\delta_\text{max}=1.3$, which is technically quasi-linear but still below the critical overdensity $\delta_\text{crit}\approx1.69$ often used in HMF formalisms. This particular value was chosen to preserve physical validity of the hybrid HMF we employ.\footnote{In order to derive this bound, we first calculate the conditional HMF $\psi(M_h,z) = dn(M_h,z|\delta,v_\text{bc}=0)/dM_h$ for $\delta\in \left[-1,2\right]$, and numerically find the values of $\delta_\text{max}$ at which the $d\psi/dM_h > 0$ for any given $M_h$ (i.e., the HMF increases, instead of decreasing, with increasing halo mass which is unphysical behavior). A value of $\delta_\text{max}=1.3$ ensures physical validity for all redshifts $z\geq6$. In the simulation, no pixels reach this threshold at $z=15$, $5$ pixels (0.0002\%) at $z=10$, and 3387 pixels (0.16\%) at $z=6$, which is a reasonable approximation expected in a semi-numerical model.\label{deltamax_footnote}}

We model star-formation to begin with the formation of Pop~III stars in metal-free mini-halos following the semi-analytic prescription of \citet{2022MNRAS.514.4433M}. In each halo, Pop~III stars are assumed to form in a single burst when the halo reaches the critical mass for star-formation $M_\text{crit}$ (described in more detail in Section~\ref{enhanced_sfe}), with a constant, fixed efficiency of $f_{\star,\threei}$ and with stellar masses distributed according to the initial mass function \citep[IMF;][]{2022MNRAS.516..841G,2025NatAs.tmp..132G}. After the first generation of stars reach the end of their lives, halos undergo a period of recovery $t_\text{delay}$ due to Pop~III supernovae during which they do not form stars. Eventually, the metal-enriched gas re-collapses, and the halos start forming  Pop~II stars. We assume that Pop~II stars form continuously with a star-formation rate density (SFRD) modelled using the sub-grid prescription:
\begin{equation}
    \begin{split}
    &\dot{\rho}_{\star,\rm II}(\mathbf{x},z)= \big(1+\delta(\mathbf{x})\big) \,\,\times\,\,
    \frac{1}{t_{\star,\twoi}H(z)^{-1}} \,\,\times \\
    & \int_{M_\text{crit}}^{M_\text{max}} \left({f_{\star,\twoi}}(\mathbf{x})f_{\twoi}(\mathbf{x},t_\text{delay})\frac{f_\text{g}}{f_\text{b}}(\mathbf{x},M_h)\dfrac{\Omega_\text{b}}{\Omega_\text{m}}M_h\right) \dfrac{dn(\mathbf{x},M_h,z)}{dM_h} dM_h
    \end{split}
    \label{eqn:sfrdII}
\end{equation}
where $\mathbf{x}$ is the spatial coordinate in the simulation box (with varying $\delta$, $v_\text{bc}$ and $M_\text{crit}$), and $(1+\delta)$ accounts for conversion from Lagrangian to Eulerian space \citep[see, e.g.][]{2011MNRAS.411..955M,2011MNRAS.418..906T,2023MNRAS.523.2587M}. Here, $t_{\star,\twoi}=0.2$ is the fraction of Hubble time over which Pop~II stars are assumed to form \citep[corresponding to the typical dynamical time for DM halos,][]{2019MNRAS.484..933P,2022ApJ...933...51R}\footnote{The dynamical (half-crossing) time for DM halo is ${t_\text{DM,dyn} = V_c/R_\text{vir}= \sqrt{3/(4\pi G\rho_\text{vir})} = \sqrt{6/(\Delta_cH(z)^2)}}$ for $\rho_\text{vir}=\Delta_c\rho_\text{crit}$. Different halo radius/mass definitions depending on the overdensity threshold ${\Delta_c=50-500}$ \citep[e.g.][]{2001A&A...367...27W} give values ${t_\text{DM,dyn}H(z) \sim 0.10-0.35}$. The Schmidt law \citep{1959ApJ...129..243S}, extended to general star-formation in proto-galaxies \citep{1997ApJ...481..703S}, yields the star-formation (or gas depletion) timescale ${t_\text{SF,dyn}=f_\star\rho_\text{gas}/\dot{\rho}_\text{star}}$. The two are closely linked via gas accretion rate onto galaxies in halos so that $t_\text{DM,dyn}\sim t_\text{SF,dyn}$ \citep[see, e.g., analytic models of][which are in agreement with simulations]{2012MNRAS.421...98D,2014MNRAS.444.2071D}.\label{tstar_footnote}}, $f_\twoi$ is \text{mass fraction of halos forming} Pop~II stars which depends on $t_\text{delay}$ \citep{2022MNRAS.514.4433M}, $f_\text{g}/f_\text{b}$ is the gas-to-baryon mass fraction affected by $v_\text{bc}$ in small halos \citep{2000ApJ...542..535G,2009MNRAS.399..369N,2011MNRAS.418..906T}, and $f_{\star,\twoi}$ is the Pop~II star-formation efficiency (the quantity of interest in this work). Since we intend to focus on Pop~II stars, Pop~III star-formation parameters are fixed to $f_{\star,\threei} = 0.002$, with a log-flat IMF in the mass range $2\msun$ to $180\msun$ \citep{2022MNRAS.516..841G}, and a $t_\text{delay}=30\myr$ recovery time \citep[e.g.][]{2018MNRAS.475.4378C,2022MNRAS.514.4433M}.

Star-formation in halos is modelled through the analytic means described above. However, the radiative fields from stellar and galactic emission propagate to cosmological distances larger than the pixel size and thus need to be handled numerically. This is done by calculating the comoving emissivity $\epsilon(\mathbf{x},\nu,z)$ of the various radiative species, and propagating them through window functions taking into account the lightcone effect and redshifting of source spectra. The sources are modelled as follows:
\begin{itemize}
    \item \textbf{X-rays}: Emitted by Pop~III and Pop~II high-mass X-ray binaries using the SEDs from \citet{2013ApJ...776L..31F}, as we shall describe in Section~\ref{theory_cxb}. We do not differentiate between the two X-ray populations here, but this distinction will be made in a future work \citep[e.g., following the prescription developed in][]{2025NatAs.tmp..132G}.
    \item \textbf{Lyman}: Lyman-$\alpha$ and Lyman-Werner bands from Pop~II stars as described in \citet{{2005ApJ...626....1B}} calculated using \textsc{Starburst99} \citep{1999ApJS..123....3L} and from Pop~III stars using the spectra from \citet{2022MNRAS.516..841G}. We account for Ly-$\alpha$ heating and multiple scattering effects \citep{2021MNRAS.506.5479R}.
    \item \textbf{Radio}: Emitted by Pop~II and Pop~III galaxies using the power-law spectrum from \citet{2018MNRAS.475.3010G, 2019MNRAS.483.1980M}, which we will describe in Section~\ref{theory_crb}. Here again, we do not differentiate radio emission between the two populations. We also exclude the effect of CMB heating \citep{2018PhRvD..98j3513V} in this work.
    \item \textbf{Ionization}: Ionization of the neutral IGM is computed using excursion set formalism \citep{2004ApJ...613....1F} with a spherical top-hat filter, where UV photons ionize the immediate surrounding of galaxies \citep{1997ApJ...475..429M,2004ApJ...609..474B}. We use the following criterion for reionization:
\begin{align}
    \exists R<R_\text{mfp}\ ,\quad {\rm s.t.}\quad \zeta_\text{ion} f_\text{coll}(\mathbf{x},R) > 1 - x_\text{e,oth}(\mathbf{x},R)
    \label{eqn:reionization}
\end{align}
where $\zeta_\text{ion}$ is the ionization rate per baryon, $R_\text{mfp}=50\cmpc$ is the maximum mean free path of ionizing photons \citep[based on end of EoR QSO observations;][]{2004Natur.432..194W,2021MNRAS.506.2390Q,2023ApJ...955..115Z}, $f_\text{coll}(\mathbf{x},R)$ is the collapse fraction of baryons into galaxies averaged over a volume of radius $R$ centered at $\mathbf{x}$, and $x_\text{e,oth}(\mathbf{x},R)$ is the fractional ionization from long-range agents like X-rays. If the criterion in Equation~\ref{eqn:reionization} is not satisfied down to cell resolution (i.e. $3\cmpc$), the cell is treated as a two-phase medium with an ionized and non-ionized fraction \citep[in agreement with cosmological radiative
transfer algorithms,][]{2011MNRAS.414..727Z,2018MNRAS.477.1549H}. 
\end{itemize}

Using the radiation fields calculated above, their effect on the gas kinetic temperature $T_\text{K}$ is calculated by solving for the thermal differential equation including X-ray heating, Lyman-$\alpha$ heating/cooling, adiabatic cooling, ionization cooling, and structure formation heating. The simulations end at $z_\text{min}=6$, close to the end of the EoR \citep[e.g.][]{2015MNRAS.447..499M,2020A&A...641A...6P,2023ApJ...942...59J}.

\begin{figure*}
	\includegraphics[width=\textwidth]{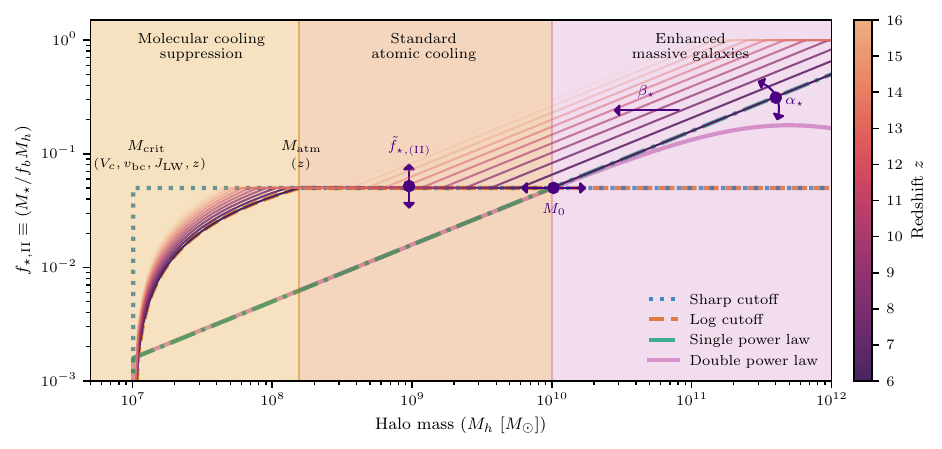}
	\caption{Schematic of the Pop~II star-formation efficiency model $f_{\star,\twoi}(M_h,z)$ used in this work (purple-brown gradient, at various redshifts), described in Equation~\ref{eqn:sfe_model}, as a function of halo mass $M_h$ and redshift $z$. The model is a four-parameter function that is zero below the critical mass $M_\text{crit}$ (set to $10^{7}\msun$ here for illustration, but can be as low as $10^{5}\msun$ at $z=30$), logarithmically increasing in the molecular cooling regime (left, light orange shaded region), constant in the atomic cooling regime (center, dark orange shaded region), and a power-law in the high mass regime (right, purple region). The turning point ${M_\text{high}=M_0\left[(1+z)/7\right]^{-\beta_\star}}$ is redshift dependent and the evolution rate $\beta_\star$ captures the enhanced star-formation efficiency at fixed halo mass in the early Universe. The SFE is naturally capped at $1$. The example illustrated here corresponds to $\tilde{f}_{\star,\twoi}=0.05, M_0=10^{10}\msun, \alpha_\star=0.5, \beta_\star=4$. Previous works using \simcode\ \citep[e.g.][]{2024MNRAS.531.1113P} utilize the log cutoff model (dashed orange line, shown at $z=6$),
    where the SFE is a fixed, constant value in the atomic cooling regime which is too simplistic for the observed mass-dependency of SFE for high mass halos. The single power law (dashed-dotted green line) on the other hand provides a better fit to large halos, but assumes a low SFE in unobserved faint, low mass halos which may not be accurate. Finally, the double power law (solid pink line) accounts for feedback from AGN in the highest mass halos, which are abundant at $z\leq6$, but their impact is less clear at high redshifts.}
	 \label{F:SMHM}
\end{figure*}

\subsection{Enhanced Pop~II star-formation efficiency}
\label{enhanced_sfe}

The star-formation prescription described in the previous section contains two important parameters: $M_\text{crit}$ and $f_{\star,\twoi}$ (for Pop~II star formation).

Halos of mass $\sim10^{5-10}\msun$ are sensitive to the threshold for star formation $M_\text{crit}$, affected by feedback from streaming velocities, Lyman-Werner photons, and UV photons. The equations governing the threshold are as follows:
\begin{align}
    & M_\text{vir} = \dfrac{R_\text{vir} V_c^2}{G} \approx 5.8 \times 10^{5}\mathrm{M}_\odot\left(\dfrac{V_c}{4.2\km\second^{-1}}\right)^{3} \left(\dfrac{1+z}{21}\right)^{-3/2};
    \label{eqn:mvir} \\
    & M_\text{mol} \approx f_{v_\text{bc}} f_\text{LW} \left(5.8 \times 10^{5}\mathrm{M}_\odot\right) \left(\dfrac{1+z}{21}\right)^{-3/2};
    \label{eqn:mcrit_mol} \\
    & M_\text{atm} \approx \left(3.5\times10^{7}\msun\right)\left(\dfrac{1+z}{21}\right)^{-3/2}.
    \label{eqn:mcrit_atm}
\end{align}
Here, the first equation gives the virial mass of a halo with circular velocity $V_c$ (the minimum required for efficient cooling of gas in halos\footnote{Note that this is simply related to a halo's virial temperature often used in the literature via:
\begin{equation}
    T_\text{vir} = \dfrac{\mu m_\text{p} V_c^2}{2k_\text{B}} \approx 635\kelvin \left(\dfrac{\mu}{0.6}\right)\left(\dfrac{V_c}{4.2\km\second^{-1}}\right)^2
    \label{eqn:tvir}
\end{equation}
where $m_\text{p}$ is the proton mass, $k_\text{B}$ is the Boltzmann constant, and $\mu$ is the mean molecular weight with $\mu=1.22$ for neutral primordial gas and $\mu=0.6$ for ionized hydrogen gas. We adopt the latter for consistency with \citet{2001PhR...349..125B} and other works, but the choice does not affect our results and simulations in any way as we directly deal with mass and velocity. It only affects our derived constraints on $T_\text{vir}$ presented in Table \ref{tab:Vc_constraints} and in Section \ref{results_Vc_constraints}.}), the second equation is the molecular cooling threshold for halos (corresponding to ${V_c\approx 4.2\km\second^{-1}}$), and the third is the atomic cooling threshold (corresponding to ${V_c\approx 16.5\km\second^{-1}}$). The factors $f_{v_\text{bc}}$ and $f_\text{LW}$ account for the effects of streaming velocity and Lyman-Werner feedback on the molecular cooling threshold, respectively, as defined in \citealt{2023MNRAS.526.4262G} \citep[see also][]{2012MNRAS.424.1335F,2013MNRAS.432.2909F,2022MNRAS.511.3657M}.

The variable $V_c$ encapsulates our uncertainty on the feedback processes in small halos (e.g. from SNe) and is a free parameter in our model, so that ${M_\text{crit}(\mathbf{x},z)=\max\left(\min\left(M_\text{mol},M_\text{atm}\right),M_\text{vir}\right)}$. During the EoR, UV photons irradiate star-forming gas in ionized bubbles, further increasing $M_\text{crit}$ (to $\sim 10^{9}\msun$) via photoheating feedback \citep[described in][accounting for ionized fractions in partially ionized cells]{2013MNRAS.432.3340S,2016MNRAS.459L..90C}.

In previous works using \simcode, Pop~II star-formation was modelled using a fixed, constant efficiency in the atomic cooling regime of DM halo masses, and a log-suppression in the molecular cooling regime as the gas cooling rate declines smoothly with virial temperature \citep{2001ApJ...548..509M,2013MNRAS.432.2909F}. This is sufficient for modelling the poorly constrained high-redshift cosmological 21-cm signal, but at lower redshifts this becomes unrealistic. There is a strong SFE -- halo mass dependence across redshifts for high mass halos, seen in both observations \citep{2010MNRAS.404.1111G,2015ApJ...799...32B,2021ApJ...922...29S} and simulations \citep{2018MNRAS.478.1694M,2018MNRAS.480.4842C,2021MNRAS.500.2127L,2022MNRAS.511.4005K,2022MNRAS.513.5621P}. Indeed, the current understanding of SFE at low redshifts ($z\lesssim6$) is that stellar feedback and AGN feedback effects lead to a suppression at the low halo mass and high halo mass end respectively, resulting in a double power-law function that peaks at $\sim10^{11-12}\msun$ \citep[e.g. \textsc{UniverseMachine};][]{2019MNRAS.488.3143B}.

In this work, we build upon the existing model in the code by introducing a new four parameter star-formation prescription, inspired by \citet{2019MNRAS.484..933P}, which is \textit{halo-mass and redshift dependent} as follows:
\begin{equation}
f_{\star,\twoi}(\mathbf{x},M_h,z) = 
\tilde{f}_{\star,\twoi} \begin{cases}
0 & M_h < M_\text{crit} \\
\dfrac{\log\left(M_h/M_\text{crit}\right)}{\log\left(M_\text{atm}/M_\text{crit}\right)} & M_\text{crit}\leq M_h < M_\text{atm} \\
1 & M_\text{atm}\leq M_h < M_\text{high}  \\
\left(\dfrac{M_h}{M_\text{high}}\right)^{\alpha_\star} & M_\text{high} \leq M_h
\end{cases}
\label{eqn:sfe_model}
\end{equation}
\noindent where the turning point is defined as 
\begin{equation}
    M_\text{high} = M_0 \left(\dfrac{1+z}{7}\right)^{-\beta_\star}.
    \label{eqn:mhigh}
\end{equation}
and $f_{\star,\twoi}$ is naturally capped at $1$. The particular form of $M_\text{high}$ anchors the SFE at $z_\text{min}=6$, where our simulations end and the UVLF data is strongest (i.e., essentially calibrating the SFE magnitude and evolution to be with respect to redshift $z=6$). Figure~\ref{F:SMHM} shows a schematic of this SFE model. Similar models with a flattening below a given mass threshold have been explored in the literature before \citep{2016MNRAS.460..417S,2024ApJ...961...50S}, the latter of whom find that a shallow-slope best fits the lensed UVLFs from HST at low halo masses, as opposed to a single power law (dash-dotted green in Figure~\ref{F:SMHM}) that assumes a steep decline. The parameterization used in our work, due to its flexibility, relaxes the assumption of declining SFE made by the single power law model for small mass halos. We also expect our model to be reasonable compared to the double power law model (solid pink line in Figure~\ref{F:SMHM}) for two reasons: (i) AGN feedback effects are generally expected for large halos $\sim10^{12}\msun$. However, few halos reach $\sim 10^{11}\msun$ at $z\geq6$ in our simulation volume of interest $\sim10^{8}\cmpc^3$. For comparison, multi-field determinations of UVLFs combining many HST/JWST surveys typically probe volumes of $\lesssim10^{6}\cmpc^3$ across redshift bins, making large halos even rarer. Since the 21-cm signal, and in general reionization, is sensitive to the cumulative effect of all halos down to $10^{5}\msun$, our relevant halo mass range is wider than other abundance matching and analytic inferences that usually consider $M_h \gtrsim10^{9}\msun$. (ii) The role of AGN at high redshifts is still quite uncertain. Although pre-JWST studies suggest a decline in AGNs at $z\gtrsim6$ \citep{2019MNRAS.488.1035K}, more recent works indicate a non-negligible faint AGN population \citep{2023ApJ...959...39H,2024ApJ...977..250F}. The contribution of these AGN to star-formation quenching remains unclear since the bright end of the luminosity function is poorly sampled at high redshifts.

The redshift evolution factor $\beta_\star$ is a timely addition, motivated by the abundance of bright galaxies in recent high-redshift JWST observations. Physically, the term captures an an enhancement in the SFE in the early Universe (for $\beta_\star>0$), meaning \textit{DM halos of the same mass host more massive and brighter galaxies in the early Universe than today}.

\subsubsection{Phenomenological comment on the SFE}

The SFE defined here is the \textit{fraction of the accreted gas in a dark matter halo that is converted into stars}, sometimes referred to as the integrated baryon conversion efficiency \citep[e.g.][]{2018MNRAS.477.1822M,2022A&A...663A..85Z}. To first order, the SFE is equal to the stellar to halo mass fraction divided by the baryon fraction $f_\text{b}\approx0.16$.

We note here that the SFE implementation in our sub-grid model is essentially an ensemble average since we do not track individual halos. For a given halo mass and redshift, the SFE of the galaxy in the host halo is fixed (i.e., there is no galaxy-to-galaxy scatter around the mean). This approach is reasonable as both the 21-cm signal and UVLF are statistical quantities probed at large scales \citep[but see also][on the importance of stochasticity]{2022MNRAS.511.5265R,2024A&A...692A.142N}. Observations of individual galaxies can significantly differ from the `global mean', and indeed any observational dataset is biased to the brightest and most massive galaxies which will be inherently stochastic \citep[e.g][]{2023MNRAS.519..843M}. Nonetheless, rare galaxies manifest in our simulations via our pixel-level formalism. The abundance of halos and their gas-to-baryon mass fraction ($f_\text{g}/f_\text{b}$) is conditional on the halo mass, and pixel-level overdensity and streaming velocity. The minimum mass for star-forming halos $M_\text{crit}$ is further affected by local feedback mechanisms. Thus, the stellar mass and UVLF still varies between pixels across the simulation. Rare overdensities at high-$z$ will produce rare high mass galaxies, which can have an elevated SFE if they fall in the power-law regime of Equation~\ref{eqn:sfe_model}. The main challenges of SFE enhancement (at any redshift) is feedback, and so there are two interpretations of this redshift evolution owing to differing star formation histories:
\begin{enumerate}[(i)]
    \item The first interpretation is one of bursty star-formation, where episodes of high-SFE can be induced by large-scale inflows \citep{2025arXiv250300106M}, positive feedback from AGN \citep{2024ApJ...961L..39S}, or delayed stellar feedback \citep[][which lead to a flattened SFE slope in the low halo mass regime; see their Figure~5]{2022MNRAS.511.3895F}. An extreme version of the last case are the ``feedback-free bursts'' (FFB) recently discussed in \citet{2023MNRAS.523.3201D} and \citet{2024A&A...690A.108L} where burstiness occurs on very short timescales of $\sim10\myr$ (i.e., multiple cycles/generations within a single half-crossing time of a virialized DM halo). This happens through rapid infall of gas via cold streams in low metallicity, high density environments, quicker than SNe/wind timescales, thereby essentially creating feedback-free conditions. The halo mass threshold above which these bursty galaxies form decreases with increasing redshift in this scenario, leading to an early time enchancement \citep[see Figure~2 of][]{2024A&A...690A.108L}.  
    \item The second interpretation is the standard uniform star-formation, through sustained gas supply and accretion to the galaxy over the halo half-crossing time. If the conditions in the early Universe are viable for compact galaxies hosting a large number of high density stellar clusters which have short dynamical timescales, they may be resilient to feedback processes essentially creating a ``feedback failure'' environment where stars can form with high efficiencies. These conditions could be due to increased binding energy (or deeper potential wells) of small halos at high-$z$ \citep{2014MNRAS.445.2545D,2017MNRAS.472.1576F}; dark-matter accelerated gas collapse \citep{2025MNRAS.538.3210B}; or a SFE-scaling that increases with gas surface density analogous to giant molecular clouds \citep[density modulated SFE or DMSFE;][]{2025arXiv250505442S}.
\end{enumerate}

\subsection{Simulating observables of interest}
\label{observables}

\subsubsection{The 21-cm signal}

The 21-cm spectral line arises from the forbidden transition of a neutral hydrogen atom from the hyperfine ground state (electron and proton spins anti-aligned) to an excited state (spins aligned). The vast abundance of neutral hydrogen at cosmic scales, however, ensures that enough atoms undergo this process creating observable signals. Whether the signal is seen in emission or absorption depends on the relative occupancy of the hyperfine levels \citep{1956ApJ...124..542P, 1990MNRAS.247..510S}:
\begin{equation}
    \dfrac{n_1}{n_0} = 3\exp\left(\dfrac{-h \nu_{21}}{k_\text{B}T_\text{S}}\right)
\end{equation}
where $n_0$ and $n_1$ are number densities of hydrogen atoms in the lower and higher energy states, the factor of $3$ accounts for the statistical degeneracy of the higher energy triplet state, ${\nu_{21} = 1420\mhz}$ is the rest-frame frequency of the 21-cm line, and $\TS$ is the spin temperature of the hydrogen atoms.

As the Universe expands, this 21-cm signal is observed today as a distortion to the CMB black body spectrum (or radio background, with temperature $T_r$) across the radio frequency range $\nu_{21,\text{obs}}=\nu_{21}/(1+z)$, corresponding to the different epochs at redshifts $z$. The strength of the signal is usually quantified in terms of the differential brightness temperature
\begin{equation}
    T_{21}(z) = \left(1-e^{-\tau_{21}}\right)\dfrac{\TS-T_r}{1+z},
    \label{t21equation}
\end{equation}
where $\tau_{21}$ is the 21-cm radiation optical depth given by
\begin{equation}
    \tau_{21}(z) \approx \dfrac{3}{32\pi} \dfrac{hc^3 A_{10}}{k_\text{B}\nu_{21}^2}\dfrac{\xhi(z) n_\text{H}(z)}{(1+z)dv_\parallel/dr_\parallel}\dfrac{1}{\TS(z)},
\end{equation}
with $x_\text{HI}$ being the fractional abundance of neutral hydrogen, $n_\text{H}$ the hydrogen number density, $A_{10}$ the spontaneous emission rate of the 21-cm transition, and $dv_\parallel/dr_\parallel$ the proper velocity gradient along the observer line of sight. If ${T_\text{S}>T_r}$, the signal is seen in net emission against the radio background, otherwise it is seen in net absorption.

In order to calculate this brightness temperature, one needs the spatially and temporally varying fractional abundance $\xhi$ and the spin temperature $\TS$. The former depends on the distribution of ionizing stellar/quasar sources, while the latter is an interplay between several competing influences that can cause a spin-flip transition in hydrogen. Coupling of $\TS$ to the background radiation temperature $T_r$ is mediated via scattering of background photons, while coupling to the kinetic temperature of matter $T_\text{K}$ is through collisions between atoms/electrons in dense IGM, and the Wouthusen-Field (WF) effect when star-formation begins \citep{1952AJ.....57R..31W,1958PIRE...46..240F}. Thus, we get the equation \citep{2006PhR...433..181F}:
\begin{equation}
    \TS(z)^{-1} = \dfrac{T_r^{-1} + x_c T_\text{K}^{-1} + x_\alpha T_\text{C}^{-1}}{1 + x_c + x_\alpha}
\end{equation}
where $x_c$ and $x_\alpha$ are the coupling strengths of the latter two processes respectively, and $T_\text{C}~(\approx\TK)$ is the colour temperature of the Ly$\alpha$ radiation field that intermediates the WF effect.

Current experiments are either aimed at detecting the sky-averaged global signal \citep[e.g.][]{2018Natur.555...67B, 2019JAI.....850004P,2022NatAs...6..607S,2022NatAs...6.1332D}:
\begin{equation}
    \langle T_{21}(z) \rangle = \int T_{21}(\mathbf{\hat{r}},z) d\Omega ,
\end{equation}
where $\mathbf{\hat{r}}$ is a line-of-sight on the sky and $\Omega$ is the solid angle; or the power spectrum $P_{21}(k,z)$ \citep[e.g.][]{2016MNRAS.460.4320E,2020MNRAS.493.1662M,2022ApJ...925..221A}:
\begin{equation}
    \bigl\langle \tilde{T}_{21}(\mathbf{k},z) \tilde{T}_{21}^*(\mathbf{k'},z) \bigr\rangle = (2\pi)^3 \delta^D(\mathbf{k}-\mathbf{k'})P_{21}(k,z)
\end{equation}
where $\tilde{T}_{21}(\mathbf{k},z)$ is the Fourier-transform of $T_{21}(\mathbf{x},z)$, $k$ is the co-moving wavevector and $\delta^D$ is the 3D Dirac delta function. In this work, we use the power spectrum in its conventional `dimensionless' form:
\begin{equation}
\Delta_{21}^2(k,z) = (k^3/2\pi^2)P_{21}(k,z).
\end{equation}
For detailed pedagogical reviews on 21-cm cosmology, we refer the readers to \citet{2001PhR...349..125B,2006PhR...433..181F, 2012RPPh...75h6901P, 2016PhR...645....1B, 2019cosm.book.....M}. The calculation of the 21-cm global signal and power spectrum is done in post-processing in \simcode.

\subsubsection{Diffuse X-ray background}
\label{theory_cxb}

X-ray emissions are another key ingredient of the early Universe, contributing to both the heating of the IGM and the ionization of the neutral hydrogen. Despite its importance, much remains uncertain regarding its properties: soft X-rays ($E\sim20\ev$) in the form of thermal emission from supernovae remnants \citep{2001ApJ...553..499O,2001ApJ...563....1V,2004MNRAS.352..547R} and mini-quasars \citep{2006ApJ...637L...1K,2010MNRAS.401.2635C}, hard X-rays ($E\sim3\kev$) from black hole X-ray binaries \citep[XRBs;][]{2011A&A...528A.149M,2013ApJ...764...41F}, and AGN have all been suggested as dominant sources of X-ray emission. The properties of the X-ray background such as its strength, inhomogeneous heating effect \citep{2007MNRAS.376.1680P,2014MNRAS.445..213F}, and shape of its spectral energy distribution \citep[SED;][]{2014Natur.506..197F,2014MNRAS.445..213F,2014MNRAS.443..678P} affect the evolution of the IGM temperature, and thus the 21-cm signal coupled to it. In addition to this, the contribution of ancient X-ray sources to the unresolved X-ray background seen today could act as a useful observable  of the early Universe \citep{2017MNRAS.464.3498F}, setting upper limits on their emission efficiency.

\begin{figure*}
	\includegraphics[width=\textwidth]{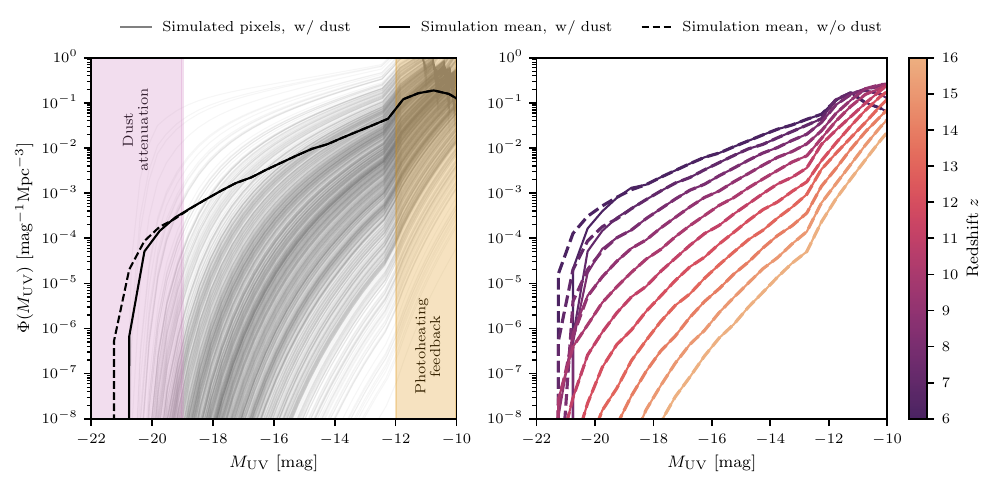}
	\caption{UV luminosity functions (UVLFs) in the simulation for a fiducial parameter set: ${V_c = 4.2\km\second^{-1}}, {\tilde{f}_{\star,\twoi}=10^{-3}}, {M_0 = 5\times10^{9}\msun}, {\alpha_\star=1.0}, {\beta_\star = 0.5}$. \textbf{Left}: UVLF at $z=8$ for a random sample of $\sim1000$ pixels (grey), averaged across the simulation volume with and without dust effects (black solid and black dash-dotted lines respectively). The suppression at the bright end is a result of dust attenuation and a finite simulation volume (left, purple shaded region), while the suppression at the faint end is due to photoheating feedback during the EoR (right, orange shaded region). The finite volume and photoheating feedback are simulation features not captured by simple analytic models. \textbf{Right}: Redshift evolution of the UVLF from $z=6$ to $z=15$ for the same parameter set.}
	 \label{F:UVLF_example}
\end{figure*}

XRBs may be a dominant source of X-ray emissions at $z\gtrsim6$, in particular HMXBs \citep[HMXBs;][]{2013ApJ...776L..31F,2016ApJ...825....7L}. A simple model for their X-ray luminosity, following local starburst-like galaxies, assumes proportionality to the halo star-formation rate \cite[SFR; e.g.][]{2003MNRAS.339..793G,2010ApJ...724..559L,2012MNRAS.419.2095M,2014MNRAS.437.1698M}:
\begin{equation}
    \dfrac{L_X}{\mathrm{SFR}} = f_X \times \left(3\times 10^{40}{\erg\second^{-1}}\msun^{-1}\yr\right)
\label{fX_equation}
\end{equation}
where $f_X$ is the X-ray emission efficiency of sources at high-redshifts, normalized to the theoretical prediction for low metallicity HMXBs \citep{2013ApJ...764...41F,2013ApJ...776L..31F,2021ApJ...907...17L} as explained in \citet{2017MNRAS.464.3498F}. The comoving X-ray emissivity throughout the simulation can then be calculated as ${\epsilon_X(\mathbf{x},E,z)= L_X/\text{SFR}\times\dot{\rho}_{\star,\twoi+\threei}(\mathbf{x},z) \times \hat{\epsilon}_X(E)}$ where $\dot{\rho}_{\star,\twoi+\threei}$ is the total Pop~II and Pop~III SFRD, and $\hat{\epsilon}_X(E)$ is the normalized X-ray SED in $\ev^{-1}$ \citep[the mean model from][at $z=15.34$ in the energy range $0.2-95\kev$, including interstellar absorption]{2013ApJ...776L..31F}. The present-day angle-averaged specific intensity from these sources at $z>z_0$ is thus \citep[e.g.][]{2007MNRAS.376.1680P}:
\begin{equation}
    J_X(E,z=0) = \frac{1}{4 \pi} \int_{z_0}^{\infty} \epsilon_X(E',z')e^{- \tau_X(E',z')} \bigg|\dfrac{cdt}{dz'}\bigg|dz',
\label{JX_equation}
\end{equation}
usually quoted in units of $\erg\second^{-1}\ev^{-1}\cm^{-2}\sr^{-1}$. In this equation, $E'=E(1+z')$ is the energy at the emission redshift $z'$, $\epsilon_X(E', z')$ is averaged across the simulation box\footnote{There is no need for shell radiative transfer here since we're interested in the sky-averaged diffuse background.}, $\tau_X(E', z')$ is the optical depth of X-rays calculated using the weighted cross section of hydrogen and helium species \citep{1996ApJ...465..487V}, and $H(z')$ is the Hubble parameter at emission. The observed integrated flux in some energy band is then given by:
\begin{equation}S_X\left(E_\text{band}\right)=\int_{E_\text{band}}J_X(E,z=0)dE
    \label{eqn:SX_equation}
\end{equation}
in units of $\erg\second^{-1}\cm^{-2}\sr^{-1}$.
In the case of our simulations, we use $z_0 = z_\text{min}=6$, which means that only high-redshift sources are included in the redshift integral. This complements the idea that the X-ray background measurements serve as an absolute upper limit on the contribution from high-$z$ sources. Note that we assume ${f_{X}=}f_{X,\twoi} = f_{X,\threei}$ in the constraints here; the largest contribution to the integral comes from lower redshifts, where Pop~II stars dominate the star-formation rate.

\subsubsection{Diffuse radio background}
\label{theory_crb}
In a similar manner to X-rays, one can estimate the radio luminosity per unit frequency from galaxies as \citep{2018MNRAS.475.3010G, 2019MNRAS.483.1980M, 2020MNRAS.499.5993R}:
\begin{equation}
    \dfrac{L_r(\nu)}{\text{SFR}} = f_r \times \left(10^{22} \watt\second^{-1} \hz^{-1}\msun^{-1}\yr\right) \left(\frac{\nu}{150\mhz}\right)^{-\alpha_r}
    \label{fr_equation}
\end{equation}
where $f_r$ is the emission efficiency with respect to the present-day radio galaxies, and the spectral index $\alpha_r=0.7$ matches the observed power spectrum of synchrotron emission from low redshift radio galaxies \citep{2016MNRAS.462.1910H, 2018MNRAS.475.3010G}. As discussed in the introduction, the excess radio background seen today \citep{2011ApJ...734....5F,2018ApJ...858L...9D}, if non-Galatic, could be due to a variety of exotic sources. The above formalism attributes this, at least in part since the background is an upper limit, to astrophysical origins in galaxies.

To estimate the contribution of high-redshift galaxies at $z>z_0$ to the present-day sky-averaged excess radio temperature, we use the equation \citep[e.g.][]{2020MNRAS.492.6086E,2020MNRAS.499.5993R}:
\begin{equation}
    T_r(\nu,z = 0) =  T_\text{CMB} + 
    \frac{c^2}{2k_{\rm B} \nu^2} \dfrac{1}{4\pi}
    \int_{z_0}^{\infty} \epsilon_r(\nu',z')\bigg|\dfrac{cdt}{dz'}\bigg|dz',
    \label{Tr_equation}
\end{equation}
where $\nu'=\nu(1+z')$ is the frequency at emission redshift $z'$, $\epsilon_r(\nu, z')$ is the comoving radio emissivity averaged across the simulation volume, ${z_0 = z_\text{min}=6}$ as before, and we assume that ${f_{r}=f_{r,\twoi} = f_{r,\threei}}$. Unlike X-rays, the radio emission is not absorbed as it propagate through space and time. We include the effect of line-of-sight fluctuations introduced recently in \simcode\ in \citet{2024MNRAS.52710975S}. We also note that we have not included the effect of soft photon heating from the excess background, which can have leading order effects on the matter temperature and thus the 21-cm signal \citep{2024MNRAS.534..738C}, and leave this for future work.

\subsection{UV luminosity function}
\label{uvlf_model}

\subsubsection{Analytic model}
\label{uvlf_model_analytic}

We now discuss the modelling of ultraviolet luminosity functions (UVLFs) used in this work, which is a new output in \simcode. UVLFs are a key statistic of galaxy surveys, which quantify the abundance of galaxies as a function of the rest-frame luminosity measured in the band $\lambda = 1500 \pm 50 \angs$. This is a useful tool to study the evolution/properties of galaxy populations. The main features of UVLFs can be reproduced within a framework of the HMFs \citep[see, e.g.,][]{2004MNRAS.353..189V,2019MNRAS.484..933P}:
\begin{equation}
    \Phi(\mathbf{x},M_\text{UV},z) = \big(1+\delta(\mathbf{x})\big)\dfrac{dn(\mathbf{x},M_h,z)}{dM_h}\left|\dfrac{dM_h}{dM_\text{UV}}\right|,
    \label{eqn:uvlf_def}
\end{equation}
where $\Phi$ is the number density of galaxies at UV magnitude $M_\text{UV}$ (defined later in Equation~\ref{eqn:luv_muv_relation}), $dn/dM_h$ is the hybrid HMF discussed in Section~\ref{overview21cmspace}, and the $(1+\delta)$ accounts for the conversion of the HMF from Lagrangian to Eulerian space as in Equation~\ref{eqn:sfrdII} \citep[e.g.][]{2011MNRAS.418..906T}. In order to map $M_h \leftrightarrow M_\text{UV}$, we assume that the SFR is proportional to the rest-frame UV luminosity of galaxies $L_\text{UV}$,
\begin{equation}
    \dot{M}_{\star,\twoi}(\mathbf{x},M_h,z) = \dfrac{ M_{\star,\twoi}(\mathbf{x},M_h,z)}{t_{\star,\twoi}H(z)^{-1}} = \kappa_\text{UV} \times L_\text{UV}(\mathbf{x},M_h,z)
\end{equation}
where $M_{\star,\twoi}$ is the term in parentheses in the integrand of Equation~\ref{eqn:sfrdII}, $t_{\star,\twoi}=0.2$ is the characteristic time for formation of Pop~II stars in units of the Hubble time, as discussed before in Section~\ref{overview21cmspace}, and $\kappa_\text{UV} = 1.15\times10^{-28}\msun\yr^{-1}\erg^{-1}\second\hz$ is a constant conversion factor. We adopt this value, following \citet{2014ARA&A..52..415M} who calculate it assuming a continuous star-formation history, a Salpeter IMF in the range $0.1\rangeto100\msun$, and an evolving metallicity of $Z_\star = 10^{-0.15z}Z_\odot$. Note that these assumptions regarding the stellar population are not necessarily justified \citep[e.g.][]{2018ApJ...868...92T,2023ApJS..265....5H} and indeed \citet{2014ARA&A..52..415M} caution about the metallicity dependence being calibrated to low redshifts only. At high redshifts, the metallicity could be drastically lower, with a different IMF leading to a much larger $\kappa_\text{UV}$. We use the value above for consistency with the rest of literature, but briefly explore its impact on the UVLF, together with those of $\delta_\text{max}$ and $t_{\star,\twoi}$, in Appendix~\ref{uvlf_variation}. We find that the difference between assuming a fixed Chabrier or Salpeter IMF is negligible compared to modelling uncertainties, although noting that more extreme IMFs like top-heavy Pop~III stars that are not explored here could generate larger differences \citep[e.g.][]{2022ApJ...938L..10I}.

The UV luminosity thus calculated is then related to the absolute magnitude (in the monochromatic AB magnitude system) through the standard relation \citep{1983ApJ...266..713O}:
\begin{equation}
    \log_{10}\left(\dfrac{L_\text{UV}}{\erg\second^{-1}\hz^{-1}}\right) = 0.4 \times (51.63 - M_\text{UV}).
    \label{eqn:luv_muv_relation}
\end{equation}
Although Equation~\ref{eqn:uvlf_def} is an analytic prescription, we stress that it is applied at a pixel level ($3\cmpc$ scale) before averaging over the simulation volume as shown in Figure~\ref{F:UVLF_example}. This simulation-based approach has two benefits over simple analytic models: (i) we can capture the `extreme statistics' of rare overdensities in finite simulation volume, and (ii) each pixel has a unique $M_\text{crit}$ from photoheating feedback during the EoR, giving rise to a natural suppresion of faint galaxies. The model can also be easily extended to simulate larger volumes and study large-scale overdensity biases, or include spatially varying metallicities which would alleviate the need for a constant $\kappa_\text{UV}$. We leave such investigations for future work.

\subsubsection{Dust correction}

Dust plays a crucial role in the evolution of a galaxy, as a key component in chemical reactions across the ISM, and also as an absorber of UV starlight. Formed from stellar outflows such as supernova ejecta or AGB winds \citep{1989IAUS..135..445G}, the dust alters the observed galactic spectrum by emitting the absorbed energy back in the infrared wavelengths \citep[e.g.][]{2001PASP..113.1449C, 2001ApJ...548..296W}. This attenuation in the UV regime is especially important at the low redshift end of our simulations, $z\sim6$ since HST provides the tightest UVLF constraints there. 

In order to correct for dust in our UVLF model, we add an attenuation term to the magnitudes: 
\begin{equation}
    M_\text{UV}^\text{dust} = M_\text{UV}^\text{dust-free} + A_\text{UV}(M_\text{UV}^\text{dust}).
    \label{eqn:dust_attenuation}
\end{equation}
A common simple method for estimating the correction term $A_\text{UV}$ is to use the UV spectral slope $\beta$ as a proxy for the dust distribution across galaxies, i.e. the flatter the slope, the redder the galaxy and the more dust there is. Based on the properties of over 50 starburst galaxies at $z\approx3$, \cite{1999ApJ...521...64M} derive the widely used infrared-excess (IRX) and $\beta$ slope relation:
\begin{equation}
    A_\text{UV} = C_0 + C_1\beta
    \label{eqn:irxb_relation}
\end{equation}
with best-fit parameters $C_0=4.33$ and $C_1=1.99$. Note that to avoid unphysical attenuation, the above relation is only valid for $\beta >-2.33$, and galaxies with larger UV slopes are assumed to have negligible dust effects.

The UV spectral slope is not constant and can strongly vary with UV magnitude as well as redshift of the galaxy \citep[e.g.][]{2014ApJ...793..115B}. We use the parametrisation of \cite{2024JCAP...09..018Z},
\begin{equation}
    \langle\beta\rangle(M_\text{UV}^\text{dust},z)= \dfrac{d\beta}{dM_\text{UV}^\text{dust}}(z)\left[M_\text{UV}^\text{dust}-M_{\text{UV},0}\right] + \beta_{M_{\text{UV},0}}(z),
    \label{eqn:beta_eqn}
\end{equation}
where the observations from HST in \cite{2014ApJ...793..115B} and JWST JADES in \cite{2024MNRAS.529.4087T} are combined to give the best-fit values 
\begin{align*}
    d\beta/dM_\text{UV}^\text{dust} &= -0.012z - 0.216, \\
    \beta_{M_{\text{UV},0}}(z)&=-0.081z-1.58
\end{align*}
for the reference magnitude $M_{\text{UV},0}=-19.5\magn$. Using an aggregate fit such as Equation~\ref{eqn:beta_eqn} avoids the issue of abrupt jumps or incompatible results from different HST and JWST observations and survey sizes.

Combining Equations~\ref{eqn:dust_attenuation}, \ref{eqn:irxb_relation} and \ref{eqn:beta_eqn}, we get an analytic dust correction model that can be applied across the whole simulation:
\begin{equation}
    \begin{split}
        M_\text{UV}^\text{dust} = \dfrac{1}{1-C_1\frac{d\beta}{dM_\text{UV}^\text{dust}}(z)}\bigg[M_\text{UV}^\text{dust-free}&  + C_0 \\ +\,  C_1 \beta_{M_{\text{UV},0}}
        & - C_1 \dfrac{d\beta}{dM_\text{UV}^\text{dust}}(z)M_{\text{UV},0}\bigg].
    \end{split}
\end{equation}

Physically, this leads to a suppression of the brightest galaxies at redshifts $z\lesssim10$, but does not significantly affect galaxies at $z\gtrsim10$. This is because galaxies are generally not large enough to reach the high magnitudes where dust effects become important, and there is an `expected' absence of dust in this calibration. A potential mechanism for lack of dust in galaxies at high-$z$ (the so-called `blue monsters') is by radiation driven outflows \citep{2023MNRAS.520.2445Z,2023ApJ...943L..27F,2025A&A...694A.286F}.

\subsection{Simulation parameters}
\label{simulation_parameters}

\begin{table}
\renewcommand\arraystretch{1.1}
 \caption{Parameter priors used in our Bayesian analysis. The astrophysical parameters consist of those concerning star-formation ($V_c, \tilde{f}_{\star,\twoi}, M_0, \alpha_\star, \beta_\star$) and others concerning radiation from stellar/post-stellar sources ($\tau, f_X, f_r$). The SARAS parameters consist of a 6th order polynomial foreground model and a thermal noise term, and are only included in the constraints that use SARAS~3 data (see Section~\ref{saras_data} for data description and \ref{ss:likelihoods} for the Bayesian likelihood that includes these nuisance parameters). The prior ranges for the SARAS parameters are chosen to be centered around the best-fit values in \citet{2022NatAs...6.1473B}.}
\label{priortable}
 \centering
 \begin{tabular}{ccccc}
  \hline
  \textbf{Type} & \textbf{Parameter} & \textbf{Prior} & \textbf{Minimum} & \textbf{Maximum} \\
  \hline 
  \parbox[t]{2mm}{\multirow{8}{*}{\rotatebox[origin=c]{90}{Astrophysical}}}
  & $V_c$ & Log-Uniform & $4.2\km\second^{-1}$ & $100\km\second^{-1}$ \\
  & $\tilde{f}_{\star,\twoi}$ & Log-Uniform & $10^{-4}$ & $10^{-0.3}$  \\
  & $M_0$ & Log-Uniform & $2\times10^{8}\msun$ & $10^{11}\msun$ \\ 
  & $\alpha_\star$ & Uniform & $0$ & $2$ \\
  & $\beta_\star$ & Uniform & $0$ & $5$ \\
   & $\tau$ & Uniform & $0.033$ & $0.075$ \\
   & $f_X$ & Log-Uniform & $10^{-3}$ & $10^3$  \\
   & $f_r$ & Log-Uniform & $10^{-1}$ & $10^5$ \\
  \hline
   \parbox[t]{2mm}{\multirow{8}{*}{\rotatebox[origin=c]{90}{SARAS}}} & $\sigma_\text{S3}$ & Log-Uniform & $0.01$\,K & $1$\,K  \\
   & $a_{\rm 0}$ & Uniform & 3.54 & 3.55 \\
   & $a_{\rm 1}$ & Uniform & -0.23 & -0.21  \\
   & $a_{\rm 2}$ & Uniform & 0 & 0.01  \\
   & $a_{\rm 3}$ & Uniform & -0.01 & 0  \\
   & $a_{\rm 4}$ & Uniform & 0 & 0.01  \\
   & $a_{\rm 5}$ & Uniform & -0.01 & 0.01 \\
   & $a_{\rm 6}$ & Uniform & -0.01 & 0.01 \\
   \hline
 \end{tabular}
\end{table}

As the early Universe remains uncertain, the cosmological and astrophysical processes described in Section~\ref{overview21cmspace} can be turned on/off or scaled through various parameters in \simcode. In this work, we focus on exploring the effect of different star-formation models. Hence, our free parameter space consists of eight parameters: $\big\{ V_c , \tilde{f}_{\star,\twoi}, M_0, \alpha_\star, \beta_\star, \tau, f_X, f_r \big\}$. The method of obtaining constraints on these parameters using Bayesian analysis closely follows that of previous works \citep{2023MNRAS.526.4262G,2024MNRAS.531.1113P,2024MNRAS.527..813B}, but with the updated SFE and inclusion of the UVLF observable. We briefly list the parameters and their prior ranges below.

\begin{itemize}
    \item $V_c$ -- Minimum circular velocity for star-formation in halos in the absence of feedback. As described in Section~\ref{enhanced_sfe}, the feedback processes increase the $M_\text{crit}$ value for each pixel, but $V_c$ sets the minimum threshold for star-formation in the absence of feedback processes ($M_\text{vir}$). We vary it in the range $V_c \in \left[4.2,100\right]\km\second^{-1}$, corresponding to $M_\text{vir} \in \left[3\times10^{6},4\times10^{10}\right]\msun$ at $z=6$ (see Equation~\ref{eqn:mvir}), and $T_\text{vir} \in \left[6.35\times10^{2},3.6\times10^{5}\right]\kelvin$ (see Equation~\ref{eqn:tvir}).
    
    \item $\tilde{f}_{\star,\twoi}$ -- Level of the Pop II star-formation efficiency. Previous works used a constant SFE in the atomic cooling regime denoted by $f_\star$ or $f_{\star,\twoi}$ \citep[e.g][]{2022NatAs...6.1473B,2024MNRAS.531.1113P}. In this work, we introduce $f_{\star,\twoi}(M_h,z)$ which can be halo-mass and redshift dependent (Equation~\ref{eqn:sfe_model}), and  define $\tilde{f}_{\star,\twoi}$ as the level of efficiency in this model. We vary this parameter in the wide range $\tilde{f}_{\star,\twoi} \in \left[10^{-4},10^{-0.3}\right]$, chosen to allow for both low and high SFE below the power-law transition. This is to ensure that we do not make any assumptions regarding faint galaxy populations that lie outside of observational constraints.
    
    \item $M_0$ -- Normalization of the turning point, $M_\text{high}$ (anchored at $z=6$), at which  the dependence of SFE on $M_h$ transitions from flat to a power-law. We vary this parameter in the range $M_0 \in \left[2\times10^{8},10^{11}\right]\msun$. The lower limit is chosen to be just above the atomic cooling threshold $M_\text{atm}(z=6) \approx 1.5\times10^8\msun$, and the upper limit is close to the maximum halo mass in the simulation volume at low redshifts. 
    
    \item $\alpha_\star$ -- The power-law index of the high mass end of the SFE, varied in the range $\alpha_\star \in \left[0,2\right]$. We choose positive values to `boost` the SFE at the high mass end in agreement with observations \citep[e.g.][]{2015ApJ...799...32B,2019MNRAS.488.3143B}. A value of $0$ corresponds to a flat SFE in the atomic cooling regime, while a value of $2$ corresponds to a steep power-law (i.e., efficiency rapidly increasing with halo mass).
    
    \item $\beta_\star$ -- Redshift evolution of $M_\text{high}$, varied in the range ${\beta_\star \in \left[0,5\right]}$. Higher values of $\beta_\star$ correspond to a faster evolution of the SFE with redshift, while a value of $0$ corresponds to no evolution (i.e., redshift-independence). This parameter is introduced to capture an effective boost in SFE in the early Universe for small halo masses, in order to explain the observed high-$z$ galaxy populations from JWST.
    
    \item $\tau$ -- In \simcode, the effective ionization efficiency of galaxies is captured by $\xi_\text{ion}$ as defined in Equation~\ref{eqn:reionization}. Since the optical depth of CMB photons to reionization $\tau$ monotonically scales with $\xi_\text{ion}$ when $V_c, f_{\star,\twoi}$ and other parameters are fixed \citep{2020MNRAS.495.4845C}, we use $\tau$ instead of $\xi_\text{ion}$ in our analysis as the free parameter. We adopt a uniform prior of $3\sigma$ around the measured \textit{Planck 2018} value of $\tau=0.054\pm0.07$ \citep{2020A&A...641A...6P}, i.e. $\tau \in \left[0.033,0.075\right]$\footnote{For late reionization scenarios, beyond the end of our simulation ($z\lesssim6$), we calculate the $\tau$ by extrapolating the ionization history as a function of redshift in log-log space.}. By indirectly using this reionization era constraint, we ensure that no unphysical reionization histories are realized in our models. We do not explore more complex parameterizations here, instead focusing on the SFE, due to uncertainties in the galaxy ionizing emissivities, escape fraction of ionizing sources, and IGM clumping which contribute to $\zeta_\text{ion}$ \citep[e.g][]{2023MNRAS.523L..35M,2024arXiv240618186D,2025MNRAS.539L..18A}.
    
    \item $f_X$ -- The X-ray emission efficiency of high-redshift galaxies, relative to the theoretical predictions of low metallicity HMXBs of \citet{2013ApJ...776L..31F}, as defined in Equation~\ref{fX_equation}. We vary this in the wide range $f_X \in \left[10^{-3},10^3\right]$. Since this parameter is degenerate with the SFR of galaxies (and thus the SFE), the wide range aptly allows for extreme parameter combinations such as very low SFE and very high $f_X$.
    
    \item $f_r$ -- The radio emission efficiency of high-redshift galaxies, relative to present-day galaxies as defined in Equation~\ref{fr_equation}. We vary this parameter in the range $f_r \in \left[10^{-1},10^5\right]$, chosen to be wide for the same reason as $f_X$.
    
\end{itemize}
The above described parameter priors are summarized in Table~\ref{priortable}. Using these, we run a suite of 30,000 simulations across the eight-dimensional astrophysical parameter space. Each simulation takes a couple of hours to run, and the entire suite takes a few weeks to complete on the DiRAC high-performance computing facility COSMA8.\footnote{See \href{https://cosma.readthedocs.io/en/latest/cosma8.html}{https://cosma.readthedocs.io/en/latest/cosma8} for more details on CPU specifications.}

\section{Observational data}
\label{observational_data}

Models created with \simcode\ are often utilized to constrain astrophysical parameters using various observational datasets, as summarized in Table~\ref{tab:other_works} (listed for ease of parsing previous literature). We build upon these works by including the UVLF as an additional observable, and by exploring the effect of the time-evolving enhancement of SFE. We describe all the datasets used in our analysis below.

\subsection{SARAS~3}
\label{saras_data}
The Shaped Antenna measurement of the background RAdio Spectrum (SARAS) experiments are a series of radio experiments designed to detect the global 21-cm signal. We use measurements of the global sky temperature from the latest experiment, SARAS~3, in the band $55\rangeto85\mhz$ \citep[$z\approx15\rangeto25$;][]{2021ExA....51..193T}\footnote{We do not include SARAS~2 measurements \citep{2017ApJ...845L..12S} in this work due to uncertainty in the systematics modelling in the dataset \citep[see][for a joint analysis of both SARAS~2 and 3]{2024MNRAS.527..813B}.}. The data consists of 15 hours of observation from a 14-day period during its deployment in Dandiganahalli Lake and Sharavati backwaters in Southern India \citep{2022NatAs...6..607S}. The data is calibrated to remove any radio-frequency interference (RFI), receiver systematics, and thermal emission from the lake. It is thus expected to consist of the global 21-cm signal, the time-averaged foreground from both Galactic and extragalactic sources, and any residual noise. 

For our analysis, since the foregrounds are expected to be smooth, we fit the data with a 6th order log-log polynomial (coefficients $a_i$) and also include a thermal noise term ($\sigma_\text{S3}$) for the residuals, as done in previous works \citep{2022NatAs...6..607S,2022NatAs...6.1473B,2024MNRAS.527..813B}. These $7+1$ parameters are varied in the prior range shown in Table~\ref{priortable}.

\subsection{HERA}
\label{hera_data}
The Hydrogen Epoch of Reionization Array (HERA) is a radio interferometer with cross-dipole feeds and the PAPER correlator (in its Phase~1), designed to detect the 21-cm power spectrum in the frequency range $100\rangeto200\mhz$ \citep{2017PASP..129d5001D}. We use the upper limits on the 21-cm power spectrum from HERA Phase 1 observations \citep{2023ApJ...945..124H}, which improve upon the previous limits from \citet{2022ApJ...925..221A}. This dataset contains 94 nights of observations using $35\rangeto41$ antennas across the experiment's entire frequency range. We use the two frequency bands least contaminated by radio frequency interference, Band~1 in Field~D ($6.25\rangeto9.25$ hours LST) and Band~2 in Field~C ($4.0\rangeto6.25$ hours LST) in the range $117.19\rangeto133.11\mhz$ and $152.25\rangeto167.97\mhz$ respectively. The lowest limits in the two bands are:
\begin{align*}
    & \Delta^2_{21,\text{Band 1}}(k=0.36\invhmpc,z\approx10.35) \leq 3496\millikelvin^2, \\
    & \Delta^2_{21,\text{Band 2}}(k=0.34\invhmpc,z\approx7.87) \leq 457\millikelvin^2.
\end{align*}
We reduce the data to only include every other $k$-bin to ensure neighbouring data points are uncorrelated in our analysis, as done in \citet{2024MNRAS.531.1113P}. As HERA employs a foreground avoidance technique, no foreground modelling is required and we marginalise residual systematics above the thermal noise using the likelihood function described in Section~\ref{emulator_training}, as done previously in \citet{2023ApJ...945..124H}.

\subsection{Cosmic X-ray background}
\label{cxb_data}
The cosmic X-ray background (CXB) is the unresolved X-ray flux from sources outside of the Milky Way Galaxy \citep[see][for a recent review]{2022hxga.book.....B}. CXB includes contribution from point sources such as AGN and galaxies that are too faint to be resolved by telescopes, diffuse emission from the hot IGM, and redshifted emission from X-ray binaries both local and distant (barring unknown instrumental systematics). Hence, measurements of the CXB should provide upper limits for the X-ray emissivity of XRBs in the early Universe \citep[e.g.][]{2017MNRAS.464.3498F} and complement the inference on star-formation efficiency (Equation~\ref{fX_equation}).

For our analysis, we use the known point-source subtracted CXB flux from Chandra X-ray Observatory \citep{2006ApJ...645...95H}:
\begin{align*}
&S_X(1-2\kev) \leq (1.04 \pm 0.14) \times 10^{-12}\erg\cm^{-2}\second^{-1}\sr^{-1}, \\
&S_X(2-8\kev) \leq (3.4 \pm 1.7) \times 10^{-12}\erg\cm^{-2}\second^{-1}\sr^{-1},
\end{align*}
and the collated data from Table 1 of \citet{2016ApJ...831..185H}:
\begin{align*}
&S_X(8-24\kev) \leq (1.832 \pm 0.042) \times 10^{-11}\erg\cm^{-2}\second^{-1}\sr^{-1}, \\
&S_X(20-50\kev) \leq (2.0 \pm 0.083) \times 10^{-11}\erg\cm^{-2}\second^{-1}\sr^{-1},
\end{align*}
which includes measurements from HEAO \citep{1980ApJ...235....4M,1999ApJ...520..124G}, BeppoSAX \citep{2007ApJ...666...86F}, INTEGRAL \citep{2007A&A...467..529C} and Swift BAT \citep{2008ApJ...689..666A}.

\subsection{Cosmic radio background}
\label{crb_data}
The cosmic radio background (CRB), aking to the CXB, is the unresolved radio flux from extragalactic sources, after the removal of Galactic foreground emission and the CMB monopole. This includes contribution from high-redshift radio galaxies (potentially AGN or Pop~III SNe), and provides upper limits for the radio emissivity of galaxies in the early Universe, alongside the joint inference on star-formation efficiency (Equation~\ref{fr_equation}).

For our analysis, we use the CRB data collated in Table 2 of \citep{2018ApJ...858L...9D}, which includes measurements from LWA1 Low Frequency Sky Survey \citep{2017MNRAS.469.4537D} in the range $40\rangeto80\mhz$, ARACDE2 \citep{2011ApJ...734....5F} in the range $3\rangeto11\ghz$, and other single frequency experiments at $22\mhz$ \citep{1999A&AS..137....7R}, $46\mhz$ \citep{1997A&AS..124..315A,1999A&AS..140..145M}, $408\mhz$ \citep{1982A&AS...47....1H,2015MNRAS.451.4311R}, and $1420\mhz$ \citep{2001A&A...376..861R}.

\begin{figure}
	\includegraphics[width=\linewidth]{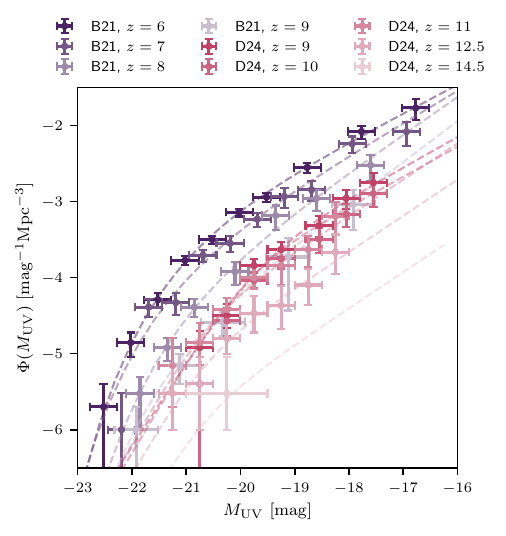}
	\caption{UVLF data used for constraints in this work. The data points are from \citet[][B21, in purple]{2021AJ....162...47B} and \citet[][D24, in pink]{2024MNRAS.533.3222D}, and represent some of the largest compilations of galaxies with HST and JWST, respectively. The dashed lines are analytic Schechter and double power law fits from the respective works.}
	 \label{F:UVLF_data}
\end{figure}

\subsection{UV luminosity function}
\label{uvlf_data}
In this work, in addition to the aforementioned datasets which have been previously used in \citet{2024MNRAS.531.1113P}, we include UVLF determinations from Hubble and JWST fields. These are:

\begin{itemize}

    \item \citealp{2021AJ....162...47B} (B21): This dataset combines several HST surveys to provides UVLFs in the range $z=2\rangeto10$. This includes the HUDF/XDF and parallel fields, BORG/HIPPIES, five CANDELS fields, HFF and parallel fields, and observations from the ERS program (see their Table~1 for details). For our analysis, we use the step-wise UVLF estimates from their Table~4 in the redshift range $z\approx6\rangeto9$ containing 1066, 601, 246 and 33 sources respectively, and omit the 9 sources at $z\approx10$ from \citet{2018ApJ...855..105O} in favour of the larger sample size in JWST data at this limiting redshift.
    
    \item \citealp{2024MNRAS.533.3222D} (D24): This multi-field dataset combines several JWST surveys, including PRIMER, NGDEEP and JADES to provide UVLFs in the redshift range $z\approx9\rangeto15$. We pick this dataset due to its inclusion of the latest JWST data, its large survey area of $\simeq 370\,\text{arcmin}^2$, and its consistent findings with other datasets at both $z=11$ \citep{2024MNRAS.527.5004M,2023ApJ...954L..46L,2023ApJ...951L...1P} and $z=12.5$ \citep{2024ApJ...965..169A,2024ApJ...970...31R}. For our analysis, we use UVLF estimates from the entire dataset given in their Table~2, consisting of a total of 2548 sources, in the redshift bins $8.5<z<9.5$, $9.5<z<10.5$, $10.5<z<11.5$, $11.5<z<13.5$ and a tentative $13.5<z<15.5$ bin based on an equivalent of $\sim1.3$ sources\footnote{ This is the total contribution from summed probability $p(z)$ of all photometric solutions; for more details on the sample selection process, see Section~3 of their work.}. However, note that due to the large error bars of the highest-$z$ data point, it does not affect our constraints strongly.

\end{itemize}

Figure~\ref{F:UVLF_data} shows the collated UVLF data used in this work. By eye, one can see that the UVLF evolution with redshift slows down at $z\gtrsim10$.

\section{Bayesian inference}
\label{inference}

\begin{table*}
    \centering
    \caption{\label{tab:emulator_table}
    Summary of neural network emulator training parameters for each observable: 21-cm global signal, 21-cm power spectrum, cosmic X-ray background, cosmic radio background, and UV luminosity function. All emulators are trained with the Adam optimization algorithm using a $\tanh$ activation function, a mean squared error loss function, and a test/train split of $1/9$. The $N_\text{samp}$ parameter denotes the number of sample points per simulation used in the emulator training (e.g. $T_{21}(z)$ consists of 250 redshift points per simulation for the global 21-cm signal emulator). The emulator accuracy quoted here is the maximum $1\sigma$ error in the part of the input space occupied by data (e.g., the SARAS~3 band $z=15-25$ for global 21-cm signal). For a complete error analysis of the emulators, see Appendix~\ref{appendix_emuaccuracy}.}
    \begin{tabular}{cccccccccc}
    \hline
    \textbf{Emulator} & \textbf{Framework} & \textbf{Network architecture} & \textbf{Batch size} & \textbf{$N_\text{samp}$} & \textbf{Emulator accuracy ($1\sigma$)} \\
    \hline
    21-cm global signal & \textsc{globalemu} & 16-16-16-16 & 250 & 250 & $<11\%$ \\
    21-cm power spectrum & \textsc{scikit-learn} (\texttt{MLPRegressor}) & 100-100-100-100 & 10000 & 20000 & $<13\%$ \\
    Cosmic X-ray background & \textsc{globalemu} & 16-16-16-16 & 400 & 400 & $<5\%$ \\
    Cosmic radio background & \textsc{globalemu} & 16-16-16-16 & 100 & 100 & $<4\%$ \\
    UV luminosity function & \textsc{scikit-learn} (\texttt{MLPRegressor}) & 100-100-100-100 & 20000 & 15000 & $<2\%$ \\
    \hline
    \end{tabular}
\end{table*}

In order to perform our parameter inference, we use a Bayesian approach similar to that of previous works \citep[e.g.][]{2022NatAs...6.1473B,2024MNRAS.531.1113P}. In this statistical paradigm, given a set of parameters $\theta$ and observed data $\mathcal{D}$, the \textit{a priori} probability of the parameters being true $\pi(\theta)$ is updated by the likelihood of the data given the parameters $\mathcal{L}(\theta) = P(\mathcal{D}|\theta)$, to give the \textit{a posteriori} probability distribution $\mathcal{P}(\theta) = P(\theta|D)$:
\begin{equation}
    \mathcal{P}(\theta) = \dfrac{\mathcal{L}(\theta)\pi(\theta)}{\mathcal{Z}},
\end{equation}
where $\mathcal{Z} = P(\mathcal{D}) = \int \mathcal{L}(\theta)\pi(\theta) d\theta$ is the Bayesian evidence (the integral of the likelihood over the parameter space). A uniform prior is uninformed and suitable for parameters which have an expected range, while a log-uniform prior works better for parameters whose order of magnitude is not known.

The Bayesian approach allows us to marginalise over the parameter space to obtain the posterior distribution of each parameter (or indeed subset of parameters) $\theta_i$ as:
\begin{equation}
    \mathcal{P}(\theta_i) = \dfrac{1}{\mathcal{Z}}\int \mathcal{L}(\theta)\pi(\theta) d\theta_{j\neq i}.
\end{equation}
Thus, we can treat the SARAS~3 foreground parameters $\theta_\text{fg}$ described in Section~\ref{saras_data} as nuisance parameters, and marginalise over them to obtain the posterior distribution of the astrophysical parameters $\theta_\text{ast}$.

We use the slice-sampling-based nested-sampling package \polychord\ \citep{2015MNRAS.450L..61H,2015MNRAS.453.4384H} to perform our Bayesian inference. A single run of \polychord\ can require millions of likelihood evaluations, which is computationally unfeasible to do by running the full \simcode\ simulation for each set of parameters. To make the inference tractable, we train neural network (NN) emulators that can rapidly evaluate the observables of interest within tens of milliseconds. We describe the emulator training and likelihood evaluation in the following sections.

\subsection{Emulator training}
\label{emulator_training}

We train emulators for our five observables of interest: the 21-cm global signal $T_{21}(z)$, the 21-cm power spectrum $\Delta_{21}^2(k,z)$, the cosmic X-ray background $S_X(E)$, cosmic radio background $T_r(\nu)$ and UV luminosity function $\Phi(M_\text{UV},z)$. We use the \textsc{Tensorflow} \citep{2016arXiv160304467A} based package \globalemu\ \citep{2021MNRAS.508.2923B} to train the one-dimensional observables: 21-cm global signal, CRB and CXB, and \textsc{scikit-learn} \citep{2011JMLR...12.2825P} to train the two-dimensional observables: 21-cm power spectrum and UVLF. All emulators are trained with the Adam optimization algorithm using a $\tanh$ activation function, a mean squared error loss function, and a test/train split of $1/9$. The input pre-processing and training details for each emulator are as follows:

\begin{itemize}
    \item \textbf{21-cm global signal}: We train the emulator on the 21-cm global signal $T_{21}$ output from the simulations in the range $z=6\rangeto30$. We pre-process the training data by first resampling the default simulation redshift step of $\Delta z=1$ to $\Delta z=0.1$ ($N_\text{samp}=250$) via simple interpolation to increase the training dataset size and resolution. We then down-scale and normalize the signals (subtracting the mean and dividing by the standard deviation of training data), as is a standard feature in \globalemu. The emulator is trained on a 4-layer NN of 16 nodes each, a batch-size of 250, and with early-stopping based on validation loss to prevent over-fitting. The resulting emulator has an accuracy of $<11\%$ error in the SARAS~3 band at 68\% confidence, and an RMSE of $\sim 70\millikelvin$ across the entire redshift range at the 95th percentile of test data. For a more detailed analysis of emulator accuracy (and its dependence on the input-space $z$), see Appendix~\ref{appendix_emuaccuracy}.
    
    \item \textbf{21-cm power spectrum}: We use a similar methodology here as described in \citet{2024MNRAS.531.1113P} using \textsc{scikit-learn}. The simulation outputs the 21-cm power spectrum $\Delta_{21}^2$ at redshifts $z=7\rangeto11$ and wavevector range $k\approx0.1\rangeto2.5\invhmpc$. We pre-process the training data by log-transforming both $z$ and $k$, as well as $\Delta_{21}^2$ (truncating values $<10^{-8}\millikelvin^2$) to improve the performance of the emulator over the large dynamic range. We then resample and interpolate the power spectrum to a finer input grid by drawing $N_\text{samp}=20000$ samples from uniform pairs of $z$ and $k$ values. The emulator is trained using an \texttt{MLPRegressor} with 4 hidden layers of 100 nodes each and a batch-size of 10000. The resulting emulator has an accuracy of $<13\%$ in linear space in the $k$ and $z$ regime of HERA data at 68\% confidence.

    \item \textbf{CXB / CRB}: We train the cosmic X-ray background $S_X$ and radio background $T_r$ in a similar manner to the 21-cm global signal. In case of the former, the CXB (Equation~\ref{eqn:SX_equation}) is output by the simulation in the energy range $E=10^{-1}\rangeto10^{3}\kev$ with $N_\text{samp}=400$. The CRB on the other hand is calculated in post-processing from star-formation rates output by the simulation (Equation~\ref{Tr_equation}) in $\nu=10\rangeto10^{3}\mhz$ with $N_\text{samp}=400$ to cover the CRB data comfortably. The data is first log-transformed in both cases, and then normalized as described above for the 21-cm global signal emulator. The emulator training parameters remain the same, except for a batch size of 400 for the CXB emulator and 100 for the CRB emulator. The resulting emulators have an accuracy of $<5\%$ and $4\%$ error in the observational energy bands and frequencies, respectively, in linear space at 68\% confidence.

    \item \textbf{UVLF}: We train the UV luminosity function $\Phi(M_\text{UV},z)$ output from the simulation in a similar manner to the 21-cm power spectrum described above. The simulation \text{code} outputs the UVLF (in log space) at integer redshifts in the range $z=6\rangeto16$ and magnitude $M_\text{UV}=\left[-23,-13\right]\magn$ with a bin size of $\Delta M_\text{UV} = 0.5$ to match most observations. The training data is preprocessed by drawing $N_\text{samp}=15000$ from uniform pairs of $M_\text{UV}$ and $z$ values, and the emulator training follows that of the 21-cm power spectrum with a batch size of $20000$. The resulting emulator has an accuracy of $<2\%$ in log space at the observed median $z$ and magnitude bins $M_\text{UV}$ at $68\%$ confidence, corresponding to an RMSE of $0.8\dex$ across the full magnitude range at 95th percentile of the test data.

\end{itemize}

The architecture and training parameters for all the emulators above have been summarized in Table~\ref{tab:emulator_table}.

\subsection{Likelihood evaluation}
\label{ss:likelihoods}

For the Bayesian inference described in Section~\ref{inference}, we evaluate the likelihood of the model given the data using the emulators described in Section~\ref{emulator_training} and the likelihood functions for each dataset we shall describe below. In this work, we introduce a new way to characterise the `model error' as the incoherent sum of the theory/simulation error and the the emulator error:
\begin{equation}
    \text{model error} = \sqrt{\text{theory error}^2+\text{emulator error}^2}.
\end{equation}
The former accounts for the assumed error on the semi-numerical model used in the simulation, fixed to an ad-hoc 20\% \citep[e.g.][]{2011MNRAS.414..727Z,2018MNRAS.477.1549H}, while the latter is derived by the performance of the emulator on test data. For a target function $f(x)$, the emulator performance may vary as a function of $x$. We evaluate this for all emulators in Appendix~\ref{appendix_emuaccuracy}, which then feeds into our likelihood functions described below.

\begin{itemize}
    \item \textbf{SARAS~3}: In order to evalulate the likelihood of the model given the SARAS~3 data, we first calculate the foreground model $T_{\rm fg}$ using the polynomial fit described in Section~\ref{saras_data}:
    \begin{equation}
        \log_{10}\left(T_\text{fg}(\nu)/\kelvin\right) = \sum_{i=0}^{6} a_i \left[f_\text{N}\left(\log_{10}\left(\nu/\mhz\right)\right)\right]^i,
    \end{equation}
    where function $f_\text{N}$ normalizes the log-frequency to be in the range $[-1,1]$. Next, we use the emulator to predict the 21-cm global signal at the frequencies of the SARAS~3 data and evaluate the likelihood function:
    \begin{equation}
        \begin{split}
            \mathcal{L}_\text{SARAS}(\theta) &= \prod_{j}^{N_\text{freq}} \frac{1}{\sqrt{2 \pi \left[\sigma_\text{S3}^2 + \sigma_{\rm model}^2(\nu_j) \right]}} \exp \bigg[\\
            &-\frac{1}{2} \bigg( \frac{T_{{\rm SARAS3},j} - T_{\rm fg}(\nu_j;\theta_\text{fg}) - T_{\rm 21,model}(\nu_j;\theta_\text{ast})}{\sigma_\text{S3}^2 + \sigma_{\rm model}^2(\nu_j)} \bigg)^2 \bigg],
        \end{split}
    \end{equation}
    where $j$ indexes the observational frequency bins, $T_{\rm SARAS3}$ is the observed SARAS~3 sky temperature corrected for known systematics, $T_{\rm fg}$ is the foreground model, $\sigma_\text{S3}$ is the thermal noise nuisance parameter as mentioned in Section~\ref{saras_data}, $T_{\rm 21,model}$ is the theory model signal, and $\sigma_{\rm model}(\nu_j)$ is the model error at the frequency $\nu_j$.
    
    \item \textbf{HERA}: We use the emulator to predict the 21-cm power spectrum at the $k$ and $z$ values of the HERA data and adopt the same likelihood function as in \citet{2022ApJ...924...51A}:
    \begin{align}
    \begin{split}
        \mathcal{L}_\text{HERA}&(\theta_\text{ast}) =\\ 
        &\prod_{j}^{N_\text{bins}} \dfrac{1}{2} \left[1 + \text{erf}\left(\dfrac{\Delta_{{\rm 21,HERA},j}^2 - \Delta_{21,\text{model}}^2(k_j,z_j;\theta_\text{ast})}{\sqrt{2\left[\sigma_{\text{HERA},j}^2 + \sigma_\text{model}^2(k_j,z_j)\right]}}\right)\right],
        \label{eq:hera_likelihood}
    \end{split}
    \end{align}
    where $j$ indexes the observational $k$ and $z$ bins, $\Delta_{\rm 21,HERA}^2$ is the observed HERA upper limits, $\Delta_{21,\text{model}}^2$ is the theory model predicted 21-cm power spectrum, $\sigma_\text{HERA}$ is the observational error, and $\sigma_\text{model}$ is the model error at those $k_j$ and $z_j$ values.

    \item \textbf{CXB / CRB}: For the CXB and CRB inference, we use a similar likelihood to HERA since we use the data in the form of upper limits. In case of the former, the observable is the CXB flux $S_X(E_\text{band})$ and we use the emulator to predict this at a high resolution within each energy band of the CXB data and then integrate the flux over the bin. In case of CRB, the observable is the temperature $T_r(\nu)$, and we use the emulator to predict the CRB flux at the precise frequencies of the CRB data. The likelihood function is then evaluated as in Equation~\ref{eq:hera_likelihood} for both datasets.

    \item \textbf{UVLF}: To evaluate the likelihood of the model given the UVLF data, we use the emulator to predict $\log_{10}(\Phi)$ at the $M_\text{UV}$ bins and median $z$ values of the UVLF observations, and use a two-piece Gaussian likelihood function to account for asymmetric data errors (as in Figure~\ref{F:UVLF_data}):

    \begin{equation}
        \begin{split}
            \mathcal{L}_{\text{UVLF}}&(\theta_\text{ast}) = \prod_{i}^{z-\text{medians}} \prod_{j}^{M_\text{UV}-\text{bins}} \sqrt{2/\pi}\left(\sigma_+ + \sigma_{-}\right)^{-1} \exp \bigg[\\
            &-\frac{1}{2} \left( \frac{\log_{10}\left(\Phi_{\text{obs},i,j}\right) - \log_{10}\left(\Phi_\text{model}(M_{\text{UV},i},z_{j};\theta_\text{ast})\right)}{\sigma_{+/-}} \right)^2 \bigg],
        \end{split}
    \end{equation}
\end{itemize}
where $\sigma_{+/-} = \sqrt{\left(\sigma_{+/-,\text{obs},i,j}\right)^2 + \left(\sigma_\text{model}(M_{\text{UV},i},z_{j})\right)^2}$ is the asymmetric error on the UVLF data point, and the denominator of the exponential is $\sigma_+$ if $\log_{10}\left(\Phi_{\text{th}}\right) > \log_{10}\left(\Phi_\text{obs}\right)$, and $\sigma_-$ otherwise.

Equipped with the likelihood function for each dataset, we can now perform our Bayesian analysis to infer the astrophysical parameters $\theta_\text{ast}$ given the data.

\begin{figure*}
    \includegraphics[width=\textwidth]{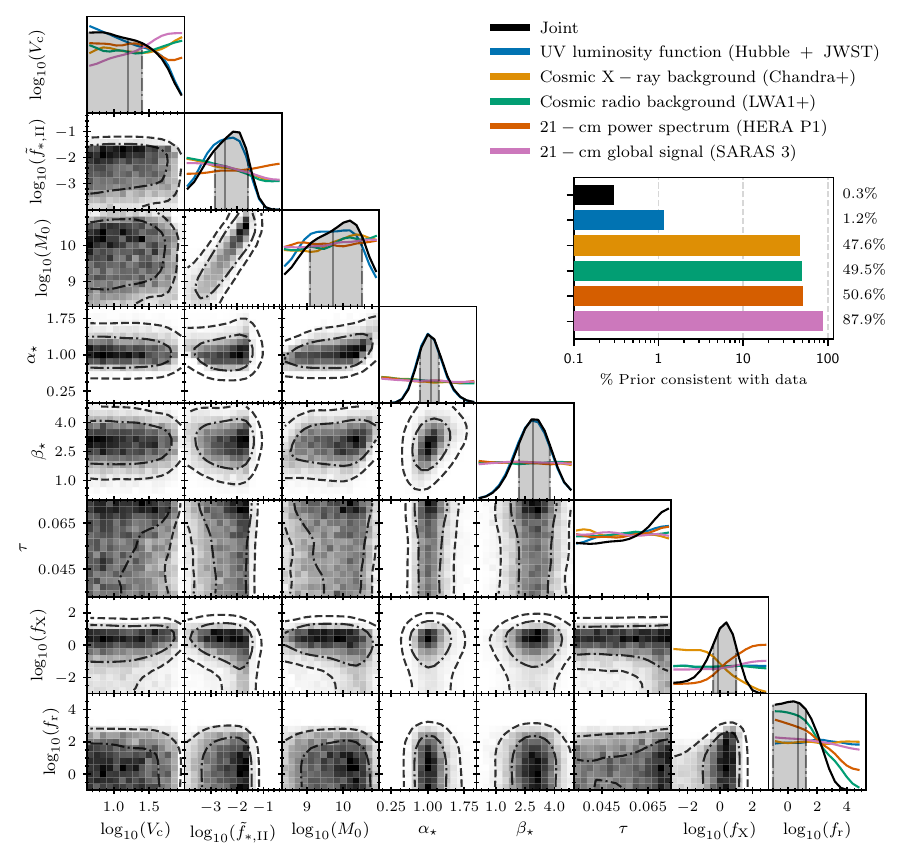}
    \caption{Marginal 1D and 2D posteriors of astrophysical parameters from the joint analysis and individual analysis of the datasets: UVLF, CXB, CRB, HERA, and SARAS~3 (color-coded as shown in the legend). The 2D posterior contours show the 68\% (dash-dotted) and 95\% (dashed) credible regions emerging from the joint analysis. The 1D marginal posterior PDFs show the weighted mean (solid line) and 68\% credible region (shaded with dash-dotted outline; calculated as the smallest CDF intervals containing the required probability volume) for the joint fit (black), as well as the 1D PDFs for each individual observable (indicated in the legend). Top right panel shows the posterior to prior volume contraction for each dataset, calculated using \textsc{margarine}. For a similar plot comparing joint constraints with and without the UVLF dataset, see Figure~\ref{F:TrianglePlot_woUVLF}, and for joint constraints with and without the SARAS~3 dataset, see Figure~\ref{F:TrianglePlot_woSARAS}. For the full parameter space, including SARAS~3 nuisance parameters, see Figure~\ref{F:TrianglePlot_Joint_full}.}
    \label{F:TrianglePlot_Joint_astro}
\end{figure*}

\begin{table*}
    \centering
    \renewcommand*{\arraystretch}{1.5}
    \caption{Marginal 1D constraints on the parameter $V_c$, and related quantities $M_\text{vir}$ (Equation~\ref{eqn:mvir}) and $T_\text{vir}$ (Equation~\ref{eqn:tvir}). The quoted errors are at 68\% (95\%) confidence levels around the weighted mean. We list SARAS~3 and UVLF individually since they are the most constraining datasets, providing lower and upper limits respectively. The joint analysis combines the two, with SARAS~3 weakening the upper bound on $V_c$ set by UVLFs as seen in Figure~\ref{F:TrianglePlot_woSARAS} (or providing a lower bound depending on the foreground fit, as in Figure~\ref{F:TrianglePlot_binsZ}).}
    \label{tab:Vc_constraints}
    \begin{tabular}{cccccc}
    \hline
    \textbf{Parameter} & \textbf{Prior} & \textbf{SARAS~3} & \textbf{UVLF} &  \textbf{Joint w/o SARAS~3} &
    \textbf{Joint} \\
    \hline
    $\log_{10}\left(V_c/\km\second^{-1}\right)$ & $\mathcal{U}\left(0.62,2\right)$ or  & $\gtrsim 1.16\, (0.72)$ & $\lesssim 1.41\, (1.79)$ & $\lesssim 1.33\, (1.78)$ & $\lesssim 1.40\, (1.79)$ \\
    $\implies V_c\ (\km\second^{-1})$ & $\log\mathcal{U}(4.2,100)$ & $\gtrsim 14.4\, (5.3)$ & $\lesssim 25.7\, (62.1)$ & $\lesssim 21.3\, (60.4)$ & $\lesssim 25.2\, (62.0)$ \\
    \hline
    $\log_{10}\left(M_\text{vir}(z=6)/\msun\right)$ & $\mathcal{U}\left(6.5,10.6\right)$ & $\gtrsim 8.09\, (6.77)$ &  $\lesssim 8.84\, (9.99)$ & $\lesssim 8.60\, (9.95)$ & $\lesssim 8.82\, (9.99)$ \\
    \hline
    $\log_{10}\left(T_\text{vir}/\kelvin\right)$ & $\mathcal{U}\left(2.8,5.6\right)$ & $\gtrsim 3.87\, (3.00)$ & $\lesssim 4.38\, (5.14)$ & $\lesssim 4.21\, (5.12)$ & $\lesssim 4.36\, (5.14)$ \\
    \hline
    \end{tabular}
\end{table*}

\subsection{Joint inference}

As mentioned previously, we use \polychord\ to perform our Bayesian inference. We sample the parameter space using 1000 live points, and use the standard stopping criterion to ensure the sampling has converged. For the joint constraints, the final likelihood is just the product of the individual likelihoods,
\begin{equation}
    \mathcal{L}_\text{Joint} = \mathcal{L}_\text{SARAS~3} \times\mathcal{L}_\text{HERA}\times \mathcal{L}_\text{CXB} \times \mathcal{L}_\text{CRB} \times \mathcal{L}_\text{UVLF} .
\end{equation}

In order to read the \polychord\ chains, generate corner-plots and perform statistics, we use the package \textsc{anesthetic} \citep{2019JOSS....4.1414H}. We also quantify how much constraining power exists in each dataset by calculating the contraction of the astrophysical prior volume to the posterior volume. In order to do this, we use the \textsc{margarine} package \citep{2023MNRAS.526.4613B} that trains masked autoregressive flows to learn marginal posterior probability density functions (PDFs; marginalizing over nuisance parameters like the SARAS~3 foregrounds). We can then calculate the Kullback-Liebler (KL) divergence statistic on the marginal posteriors to calculate:
\begin{equation}
    \begin{split}
        \text{\% prior volume consistent with data} &=100 \times \dfrac{V_\text{posterior}}{V_\text{prior}} \\ &\approx 100 \times \exp(-\mathcal{D}_\text{KL}).
    \end{split}
\end{equation}
~\\~\\
Expectedly, the KL divergence is zero when the prior and posterior are the same, and increases as the posterior diverges from the prior. We would also intuitively expect the most informative data set to be the joint constraints, which we quantify in the results section below.

\section{Results and discussion}
\label{results}

Figure~\ref{F:TrianglePlot_Joint_astro} shows the posterior on the astrophysical parameters recovered for the different datasets independently and when they are combined in a joint analysis. A detailed comparison of constraints between the datasets, excluding UVLF observations, is done in \citet{2024MNRAS.531.1113P} using an SFE model that is constant in the atomic cooling regime (dashed orange line in Figure~\ref{F:SMHM}). Their results are retrieved in our analysis as a special case of our more general SFE model --- i.e., the posterior PDF of $M_0$ saturates to the upper prior limit $10^{11}\msun$, rendering $\alpha_\star$ and $\beta_\star$ parameters redundant (since most halos are $<10^{11}\msun$ at $z\geq6$). Furthermore, the $\tilde{f}_{\star,\twoi}$ posterior PDF saturates to the lower prior limit at $10^{-4}$. This is shown in Figure~\ref{F:TrianglePlot_woUVLF}, where we compare the joint constraints with and without (shortened to w/o) the contribution of the UVLF dataset. 

The individual datasets provide different constraining power on the astrophysical parameters; the UVLF data is the most constraining, shrinking the posterior volume to $1.2\%$ of the prior volume, followed by CXB, CRB, and HERA at $\sim 50\%$ and lastly SARAS~3 at $\sim 88\%$. Note that these are individual constraints from each dataset on the full set of astrophysical parameters. Different datasets can provide more or less constraining power for specific parameters. The combination of SARAS~3, HERA, CXB and CRB show great synergies with the posterior volume corresponding to $\sim 29\%$ of the prior. With the inclusion of UVLF data, the complete joint constraints are the strongest, as expected, bringing the posterior volume further down to $0.3\%$ of the prior.

We shall present our results in the order shown in the Figure~\ref{F:TrianglePlot_Joint_astro}. We first discuss our results for the minimum circular virial velocity for star-formation in halos $V_c$ in Section~\ref{results_Vc_constraints}. We then discuss the constraints on the star-formation efficiency of galaxies in the early Universe in Section~\ref{results_sfe_constraints}, followed by the constraints on the radiative efficiencies of galaxies $f_X$ and $f_r$ in Section~\ref{results_fX_fr_constraints}.

\subsection{Constraints on $V_c$}
\label{results_Vc_constraints}

\begin{figure*}
    \centering
    \includegraphics[width=0.66\linewidth]{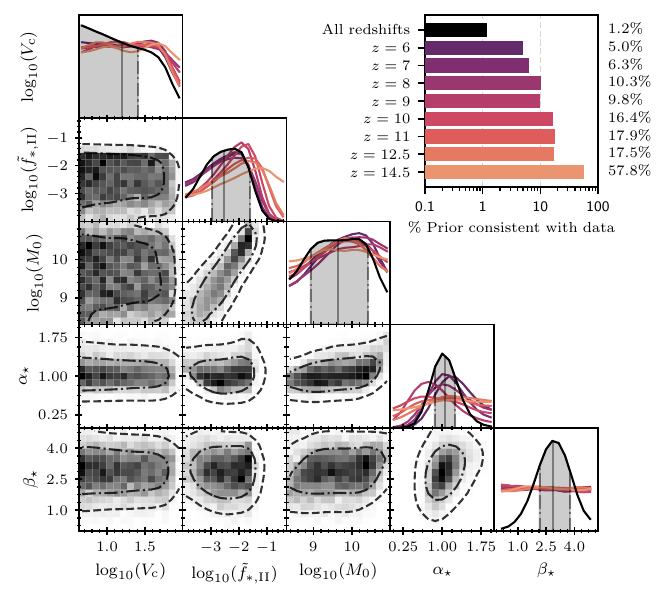}
    \caption{Marginal 1D and 2D posteriors of SFE parameters from the full UVLF dataset and individual redshift bins. As expected, the lowest redshift bins are most constraining. Furthermore, since a single redshift bin contains no redshift-evolution information, the PDF of the $\beta_\star$ parameter remains flat, and the $\tilde{f}_{\star,\twoi}$ parameter compensates by shifting the peak of the 1D posterior PDF to higher values with increasing redshift in order to explain the high-$z$ JWST observations.}
      \label{F:TrianglePlot_eachZ}
\end{figure*}

The minimum circular virial velocity $V_c$ for star-formation in halos is largely informed by the combination of SARAS~3 and UVLF data. The CXB and CRB datasets return flat 1D posterior PDFs, while HERA data has a weak (statistically insignificant) preference for lower values of $V_c$.

The dominant constraint on $V_c$ comes from the UVLFs. They directly probe faint magnitude galaxies at ${M_\text{UV} \lesssim -17\magn}$, providing a hard upper limit on how high $V_c$ can be. The data disfavour\footnote{The confidence levels (CLs) here, and for the rest of the results, are defined as the smallest intervals in the cumulative distribution function (CDF) containing the required probability volume.}:
\begin{align*}
    \log_{10}\left(V_c/\km\second^{-1}\right) &\lesssim 1.41\,(1.79) \text{ at 68\% (95\%) CL} \\
    \implies V_c &\lesssim 25.7\,(62.1) \km\second^{-1} \text{ at 68\% (95\%) CL}.
\end{align*}
Furthermore, $V_c>95\km\second^{-1}$ lies outside the $5\sigma$ confidence region, which corresponds to a DM halo mass of $M_h \approx 3\times 10^{10}\msun$ at $z=6$, and stellar masses of $M_\star \sim 10^7\msun$ in galaxies (assuming percent-level star-formation efficiency) which is consistent with the faintest galaxies observed at $z\sim6$ \citep[see, e.g., Figure~5 of][]{2024MNRAS.535.2998S}. This constraint is robust across different redshift data, as seen in Figure~\ref{F:TrianglePlot_eachZ} (first column). In order to further tighten this derived constraint, future works will include lensed luminosity functions such as the ones in \citet{2022ApJ...940...55B} that go down to $M_\star \sim 10^{5-6}\msun$ at $z\sim6$ \citep{2021AJ....162..255B}. Since our model has the added benefit of including photoheating feedback at the faint end, these lensed UVLFs can shed light on the suppression of star-formation in small halos.

Recently, \citet{2024ApJ...961...50S} performed an analysis of lensed UVLF data using truncated SFE models and found that non-lensed UVLFs disfavour a truncation of star-formation above $M_h \sim 3\times10^{10}\msun$ at $z=6$ (see their Figure~11), consistent with our results. Furthermore, they suggest that lensed UVLFs disfavour a truncation above $2\times10^{9} \msun$ at $z=6$ (i.e., strong photoheating feedback disfavoured) at $2\sigma$ confidence. This would correspond to a $V_c \lesssim 36.5\km\second^{-1}$ being favoured at $2\sigma$, indeed stronger than our $V_c\lesssim 62.1\km\second^{-1}$ constraint. Another recent multi-wavelength analysis by \citet{2025arXiv250203525Z} similar to ours also finds that UVLFs play a crucial role in providing an upper bound on the minimum halo mass required for star-formation.

We extend the aforementioned analyses by incorporating SARAS~3 data in our work, which provides a weak lower bound on the $V_c$ (or equivalently, $M_\text{vir}$) parameter space. The SARAS~3 data residual has an RMS of order $200\millikelvin$ after foreground modelling and removal, essentially setting an upper limit on the amplitude of the 21-cm absorption signal in the $z=15\rangeto25$ redshift band. Hence, it would favour 21-cm signals that either have shallower absorption troughs or have the deepest absorption at redshifts lower than $z\lesssim15$. Physically, the latter implies a delayed contribution of first stars and galaxies to the Ly$\alpha$ coupling, along with heating of the IGM from XRBs. In other words, star-formation is suppressed in small halos which corresponds to a higher $V_c$ threshold. This is reflected in the 1D posterior PDF, which gives us:
\begin{align*}
    \log_{10}\left(V_c/\km\second^{-1}\right) &\gtrsim 1.16\,(0.72) \text{ at 68\% (95\%) CL} \\
    \implies V_c &\gtrsim 14.4\,(5.3) \km\second^{-1} \text{ at 68\% (95\%) CL}.
\end{align*}
At $1\sigma$, this $V_c$ constraint alone would imply that star-formation is suppressed in \text{most} molecular-cooling halos $(V_c < 16.5\km\second^{-1})$. At $2\sigma$ however, the data only weakly disfavours small halos, allowing for $V_c \gtrsim 5.3\km\second^{-1}$, and 1D posterior PDF does not go to zero probability at the lower end. Although this has potential to be a powerful constraint, it has two added complexities: (i) we find that $V_c$ is degenerate with the SARAS~3 foreground parameters. In particular, the lowest order polynomial coefficients $a_0$ and $a_1$ (which have the largest effect due to the nature of the log polynomial function used) show a negative correlation with $V_c$. This is because (for all other astrophysical parameters fixed) lower values of $V_c$ correspond to deeper 21-cm absorption troughs in the SARAS~3 band. The foreground model compensates for this by assuming a larger $a_0$ and $a_1$, which boosts the foreground by tens of $\millikelvin$, especially at the low redshift edge of the SARAS~3 band ($z \sim 15$). This is seen in the 2D posterior PDFs in Figure~\ref{F:TrianglePlot_Joint_full}. (ii) The SARAS~3 foreground model likely subsumes some of the 21-cm signal. Thus, in future works, it would be good to check if the choice of the form of foreground model biases our results.

\begin{figure*}
    \includegraphics[width=1\linewidth]{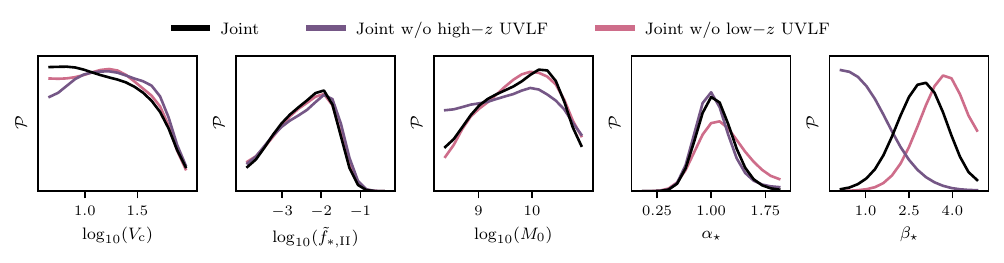}
    \caption{Marginal 1D posteriors of the SFE parameters $\tilde{f}_{\star,\twoi}$, $M_0$, $\alpha_\star$, and $\beta_\star$ from all observational datasets, with the UVLFs split into two redshift bins: $z=7\rangeto10$ as the `low-$z$' (purple) and $z=11\rangeto14.5$ as the `high-$z$' (pink) fits, with the $z=6$ data points from B21 included in both (see Footnote~\ref{z6anchoring}). The main takeaway here is that the redshift-evolution term $\beta_\star$ showing little to no evolution at low-$z$ (i.e., consistent with 0) and a rapid evolution at high-$z$. The $V_c$ constraints, discussed in Section~\ref{results_Vc_constraints}, show slight variation depending on the SARAS~3 foreground parameter, with the low-$z$ fit exhibiting the strongest synergy between the UVLF and SARAS~3 data.}
    \label{F:TrianglePlot_binsZ}
    \vspace{-0.3cm}
\end{figure*}

In the joint analysis combining all datasets, the inclusion of SARAS~3 data weakens the upper limit on $V_c$ set by UVLFs as shown in Figure~\ref{F:TrianglePlot_woSARAS}. The joint constraints yield a $V_c$ value of:
\begin{align*}
    V_c &\lesssim 21.3\,(60.4) \km\second^{-1} \text{ at 68\% (95\%) CL, without SARAS~3, and} \\
    V_c &\lesssim 25.2\,(62.0) \km\second^{-1} \text{ at 68\% (95\%) CL, with SARAS~3.}
\end{align*}
Table~\ref{tab:Vc_constraints} summarizes these constraints at both 68\% and 95\% confidence levels, converted to $T_\text{vir}$ and $M_\text{vir}(z=6)$ for ease of comparison with works that use alternate parameterizations. Given the SARAS~3 constraint's sensitivity to the foreground fit, even a marginal change in the foreground model can lead to stronger exclusion of low $V_c$ values. This is seen in Figure~\ref{F:TrianglePlot_binsZ} (first panel) in the case of joint constraints without high-$z$ UVLF data. In the absence of a more physically motivated foreground model, we refrain from over-interpreting this result but emphasize the potential for future 21-cm signal measurements in providing constraints on $V_c$.

\textit{Our results demonstrate the synergy that exists between the 21-cm signal data and high-$z$ galaxy surveys from HST/JWST, even when the former is a non-detection, in constraining the star-forming properties of small halos.}

\subsection{Constraints on the star-formation efficiency}
\label{results_sfe_constraints}

The 21-cm signal data is not informative enough yet to constrain the SFE parameters independently and, in the absence of the UVLF data, the 1D posterior PDF for the scaling of the SFE $\tilde{f}_{\star,\twoi}$ saturates to the lower limit of the prior at $10^{-4}$, while the probability of the turning point for the SFE power law slope $M_0$ saturates to the upper limit $\sim10^{11}\msun$ (see Figure~\ref{F:TrianglePlot_woUVLF}). This is consistent with predictions from a fixed constant SFE model as adopted in \citet{2024MNRAS.531.1113P}. Thus, constraints on the star-formation efficiency parameters $\{\tilde{f}_{\star,\twoi}, M_0, \alpha_\star, \beta_\star\}$ are almost entirely driven by UVLF data. Table~\ref{tab:SFE_constraints} summarizes our results at 68\% and 95\% confidence levels. As seen in the 2D posterior plot for $\tilde{f}_{\star,\twoi}$ and $M_0$ in Figure~\ref{F:TrianglePlot_Joint_astro}, there is a strong degeneracy below ${M_0} \lesssim 10^{10.5}\msun$ which is well-described by a simple linear fit to the 2D histogram:
\begin{equation}
    \log_{10}\left(M_0/\msun\right) \approx 0.9\times\log_{10}\left(\tilde{f}_{\star,\twoi}\right) + 12,
    \label{eq:M0_fstar_fit}
\end{equation}
before saturating at $\tilde{f}_{\star,\twoi} < 0.1$. This is as expected from the observational limit on faint galaxies, as discussed in the previous section. To better understand the constraining power of the UVLF data, we now perform different redshift cuts and analyze the SFE parameters inferred. 

\begin{table*}
    \centering
    \renewcommand*{\arraystretch}{1.5}
    \caption{Marginal 1D posteriors of the SFE parameters as seen in Figure~\ref{F:TrianglePlot_binsZ} above. The quoted errors are at 68\% (95\%) confidence levels around the weighted mean.}
    \label{tab:SFE_constraints}
    \begin{tabular}{ccccc}
    \hline
    \textbf{Parameter} & \textbf{Prior} & \textbf{Joint w/o high-$z$ UVLF} & \textbf{Joint w/o low-$z$ UVLF} & \textbf{Joint} \\
    & & (low-$z$ fit) & (high-$z$ fit) & (joint fit) \\


    \hline
    $\log_{10}\left(\tilde{f}_{\star,\twoi}\right)$ & 
    $\mathcal{U}\left(-4,-0.3\right)$ &
    $-2.4^{+0.9}_{-0.4}\left(^{+1.0}_{-1.2}\right)$ &
    $-2.5^{+0.9}_{-0.4}\left(^{+1.0}_{-1.2}\right)$ &
    $-2.4^{+0.9}_{-0.4}\left(^{+1.0}_{-1.2}\right)$ \\
    \hline
    $\log_{10}\left(M_0/\msun\right)$ &
    $\mathcal{U}\left(8.3,11\right)$ &
    $9.6^{+0.8}_{-0.8}\left(^{+1.2}_{-1.3}\right)$ &
    $9.8^{+0.9}_{-0.5}\left(^{+1.1}_{-1.2}\right)$ &
    $9.7^{+0.8}_{-0.6}\left(^{+1.1}_{-1.2}\right)$ \\
    \hline
    $\alpha_\star$ &
    $\mathcal{U}\left(0,2\right)$ &
    $1.0^{+0.1}_{-0.2}\left(^{+0.5}_{-0.4}\right)$ &
    $1.2^{+0.2}_{-0.4}\left(^{+0.7}_{-0.5}\right)$ &
    $1.1^{+0.2}_{-0.2}\left(^{+0.4}_{-0.4}\right)$ \\
    \hline
    $\beta_\star$ &
    $\mathcal{U}\left(0,5\right)$ &
    $\lesssim 1.4\,(2.8)$ &
    $3.6^{+0.7}_{-0.7}\left(^{+1.4}_{-1.3}\right)$ &
    $2.9^{+0.9}_{-0.7}\left(^{+1.5}_{-1.6}\right)$ \\
    \hline
    \end{tabular}
\end{table*}

\subsubsection{How much do individual redshifts contribute?}

We first quantify the constraining power of the UVLF data from different redshift bins individually. This is shown in Figure~\ref{F:TrianglePlot_eachZ}. The SFE parameter space is constrained to $\lesssim 10\%$ of the prior volume by data at $z \approx 6\rangeto9$, $\lesssim 20\%$ at $z\approx10\rangeto13$, and $\sim 60\%$ at $z\approx14.5$ as the data becomes increasingly uncertain at high redshifts.

The constraint on $V_c$ is consistent across the redshift bins. In particular, observations of faint galaxies at $z=12.5$ provide just as good constraints as those at $z=6$. Deeper observations in fainter magnitude bins, including lensed objects with JWST, are expected to provide more stringent upper limits on $V_c$ in the near future.

Similarly, the constraints on the halo-mass dependence of the SFE, $\alpha_\star$, also show consistency between redshift bins. The scatter around the mean value inferred from the combined datasets is well within errors. Furthermore, since there is no redshift-evolution information encoded in individual redshift bins, the PDF of $\beta_\star$ is flat across all redshifts. To compensate for this lack of evolution, the peak of the PDF for $\tilde{f}_{\star,\twoi}$ shifts to higher values with increasing redshift, thus hinting at an enhanced SFE at early times.

Combining all redshifts gives us a $\sim1\%$ level volume consistency of prior and data, demonstrating the strength of using UVLF data from HST and JWST across a wide redshift range. This not only provides a tighter 1D posterior on $\alpha_\star$, but also encodes important information about the evolution of the SFE. We explore this evolution in the next section.

\subsubsection{How does the SFE evolve with redshift?}

In order to properly quantify the evolution of the SFE  with redshift, $f_{\star,\twoi}\propto \left[(1+z)/7\right]^{-\beta_\star}$, we now take a closer look at the $\beta_\star$ parameter. We split the UVLF dataset into two redshift bins, now including the rest of the observational datasets (SARAS~3, HERA, CXB and CRB data) for both, to compare to the joint constraint which uses the complete UVLF dataset across all redshifts. We denote $z=7\rangeto10$ as the `low-$z$' cut and $z=11\rangeto14.5$ as the `high-$z$' cut, including the $z=6$ data point from B21 in both.\footnote{\label{z6anchoring}This `anchoring' at $z=6$ is done for two reasons. First, the SFE as defined in Equation~\ref{eqn:sfe_model} and \ref{eqn:mhigh}, is normalized to this redshift which means the $\beta_\star$ parameter has no effect at $z=6$. Secondly, by fixing $\tilde{f}_{\star,\twoi}$, $\alpha_\star$ and $M_0$ to the best-fit values at $z=6$, we can isolate the effect of SFE evolution via $\beta_\star$.} The choice of redshift bins defines the high-redshift frontier before and after the launch of the JWST, hence informing us on how the new instrument has changed our understanding of the SFE. Figure~\ref{F:TrianglePlot_binsZ} shows the result of this analysis. The $\tilde{f}_{\star,\twoi}$ and $\alpha_\star$ parameters are consistent across the two redshift ranges and $M_0$ remains degenerate with $\tilde{f}_{\star,\twoi}$, as suggested by Equation~\ref{eq:M0_fstar_fit}. However, $\beta_\star$ shows a stark contrast:
\begin{itemize}
    \item \textbf{Joint w/o high-$z$ UVLF} (or low-$z$ fit for short): At $z\leq10$, the UVLF data favours a very slow/non-evolving SFE. The 1D posterior PDF of $\beta_\star$ saturates at the lower edge of the prior, favouring $\lesssim 1.4$ at 68\% CL, and remains consistent with $\sim0$. This behaviour aligns with pre-JWST inferences using HST data \citep{2013ApJ...768L..37T,2015ApJ...813...21M,2015ApJ...803...34B,2018ApJ...868...92T,2018PASJ...70S..11H,2021AJ....162...47B,2021ApJ...922...29S}, where a simple redshift-indepenent parameterization calibrated at $z\lesssim4$ can reproduce the observed UVLFs at higher redshifts. This trend of slow evolution is also consistent up to $z\approx10$ with the FirstLight numerical simulation suite \citep{2024A&A...689A.244C}.
    
    \item \textbf{Joint w/o low-$z$ UVLF} (or high-$z$ fit for short): At $z>10$, the UVLF data favours a rapidly evolving SFE with the 1D posterior PDF of $\beta_\star$ having a weighted mean of $\langle \beta_\star \rangle = 3.6$. This deviation from the low-$z$ behaviour is an interesting outcome of the abundance of bright galaxies observed in the JWST data. The bright excess has been alternatively attributed in the literature to either a top-heavy IMF of the early populations of stars \citep{2022ApJ...938L..10I,2023ApJ...954L..48W,2024MNRAS.529.3563T,2025A&A...694A.254H}, or stochasticity/variability in star-forming galaxies \citep{2023MNRAS.521..497M,2023MNRAS.519..843M,2023MNRAS.526.2665S,2023ApJ...955L..35S,2023A&A...677L...4P,2023MNRAS.525.3254S}.\footnote{For completeness, we note that other solutions have also been proposed. This includes the presence of a hidden population of AGN \citep{2024JCAP...08..025H}, or simply that a formation epoch of $z\approx15$ implies young, bright stellar population \citep[requiring no new physics to explain current observations;][]{2025MNRAS.539.2409D}.} However, both these solutions give rise to other problems --- for example, \citet{2024A&A...686A.138C} find that a top-heavy IMF alone leads to strong stellar feedback which inhibits stellar mass growth. Similarly, \citet{2024A&A...692A.142N} and \citet{2024MNRAS.527.5929Y} conclude that larger than physical scatter would be needed to explain the bright galaxies. Although a redshift dependence in the SFE has been previously proposed \citep[e.g.][]{2016MNRAS.460..417S}, we constrain this SFE evolution for the first time at high-$z$ using a sub-grid analytic prescription. The redshift evolution is quantitatively similar to the feedback free burst (FFB) galaxies explored in \citet{2023MNRAS.523.3201D,2024A&A...690A.108L}, as well as the density-modulated SFEs (DMSFE) in \citet{2025arXiv250505442S}. Indeed, we shall see how our inferred SFE posteriors align with the former in the next section. A similar transition from low SFE to high SFE at $z\sim10$ has also been inferred in \citet[][see their Figure~10]{2025arXiv250307590C}.
    
    \item \textbf{Joint} (or joint fit): The joint constraints, combining both low-$z$ and high-$z$ UVLF data, give a weighted mean of $\langle \beta_\star \rangle = 2.9$. This fit is effectively a compromise between the two redshift bins, showing a moderate evolution of the SFE. {Although} not an ideal fit at the lower redshifts due to small error-bars on the data, it provides a simple parameterization that can be applied across the entire redshift range.
\end{itemize}
Table~\ref{tab:SFE_constraints} summarizes the constraints on the SFE parameters at 68\% and 95\% confidence levels for the different UVLF cuts. Equipped with this information, we can now look at the inferred SFE in the two redshift bins.

\begin{figure*}
	\includegraphics[width=\textwidth]{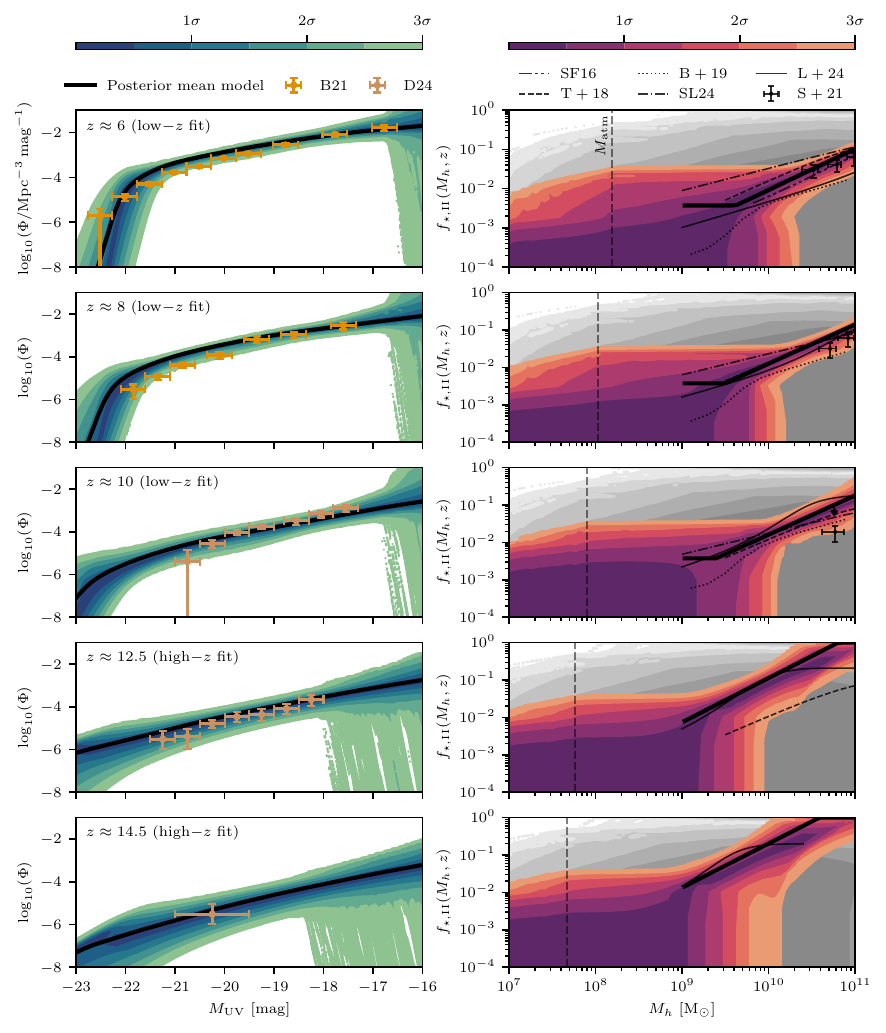}
	\caption{Functional posteriors of the UVLF (left) and constrained SFE (right) for the fit using UVLFs at low and high redshift (including all other datasets). In the upper three panels, $z=6,8,10$ we use the low-$z$ fit, while in the lower two panels, $z=12.5,14.5$ we use the high-$z$ fit (see Figure~\ref{F:TrianglePlot_binsZ} and Table~\ref{tab:SFE_constraints}). In all the panels, the filled coloured contours show $1,2,3\sigma$ confidence of the functional posterior (also showing functional SFE priors in the right panels in grey contours), while the thick black line shows the posterior-mean model (i.e., the weighted means listed in Table~\ref{tab:SFE_constraints}). Note that the black line and contours need not align since $\langle f(x|\theta_\text{ast})\rangle \neq f(x|\langle\theta_\text{ast}\rangle)$. The UVLF data (scatter points, left panels) used for the constraints are shown as a cross-check to see whether the model fits the data. In the right column, we compare the SFE posteriors to various models in the literature: halo-abundance matching results from \citet[][SF16]{2016MNRAS.460..417S}, empirical fits from \citet[][T+18]{2018ApJ...868...92T}, UniverseMachine \citep[][B+19]{2019MNRAS.488.3143B}, semi-empirical model of \citet[][SL24]{2024ApJ...961...50S}, feedback-free burst model of \citet[][L+24]{2023MNRAS.523.3201D,2024A&A...690A.108L}, and HST/Spitzer based results from \citet[][S+21]{2021ApJ...922...29S}.}
	 \label{F:SFE_UVLF_constraints}
\end{figure*}
    
\begin{figure*}
    \includegraphics[width=\linewidth]{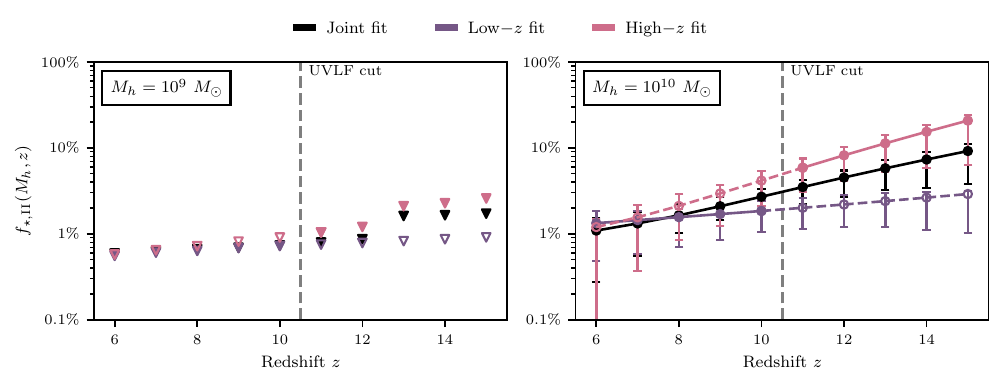}
    \caption{Evolution of SFE at fixed halo masses $M_h = 10^9\msun$ (left, triangles) and $M_h = 10^{10}\msun$ (right, circles) with redshift, for the case of the three joint analyses: low-$z$ UVLFs (purple) and high-$z$ UVLFs (pink), separated by a cut at $z\approx10.5$ (dashed vertical line), and all UVLFs (black). Unfilled markers and dashed coloured lines denote extrapolations outside of the redshift range of the fits. In the case of $M_h = 10^{10}\msun$, the error bars are $1\sigma$ confidence levels while for $M_h = 10^9\msun$, the SFE are $1\sigma$ upper limits and relatively unconstrained due to the lack of UVLF data at the faint magnitude end. See Figure~\ref{F:SFE_UVLF_constraints} for the full SFE posteriors.}
    \label{F:SFE_z_dependence}
\end{figure*}

\begin{figure}
    \includegraphics[width=\linewidth]{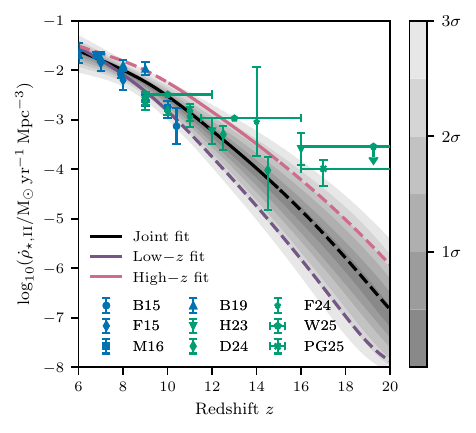}
    \caption{Evolution of the SFRD, calculated by integrating the UVLF (${\dot{\rho}_{\star,\twoi}=\int \Phi \dot{M}_{\star,\twoi} dM_\text{UV}}$) in the magnitude range $M_\text{UV} \in [-23,-17]$. The filled contours show functional posteriors of the joint fit, while the lines show the posterior-mean model (see Table~\ref{tab:SFE_constraints}) for the joint fit (black), low-$z$ fit (purple) and high-$z$ fit (pink). The lines are solid in the redshift range of the fit and dashed outside where the model is extrapolated. Various literature values are shown for comparison: results from HST in blue \citep{2015ApJ...803...34B,2015ApJ...810...71F,2016MNRAS.459.3812M,2019MNRAS.486.3805B} and JWST in green \citep{2023ApJS..265....5H,2024MNRAS.533.3222D,2024ApJ...969L...2F,2025arXiv250100984W,2025arXiv250315594P}.}
     \label{F:SFRD_constraints}
\end{figure}

\subsubsection{SFE functional posteriors and best fit}

As suggested in the previous section, the star-formation efficiency appears to have a complex evolution with time. Namely, the $\beta_\star$ parameter alone is not enough to capture the evolution at both low-$z$ and high-$z$. Thus, our inference across the full redshift range $z=6\rangeto14.5$ is essentially a composite of the low-$z$ fit at $z\leq10$ and the high-$z$ fit at $z>10$. In order to visualize the results from our Bayesian analysis  in the two redshift bins, we use \textsc{fgivenx} \citep{2018JOSS....3..849H} to convert our parameter posterior samples into functional posteriors of the SFE and its associated UVLF (as a cross-check to see how our model fits the data). This is shown in Figure~\ref{F:SFE_UVLF_constraints}. 

We also plot the posterior-mean model in the two redshift bins (Table~\ref{tab:SFE_constraints} using Equation~\ref{eqn:sfe_model}) for comparison, alongside other empirical SFE models from literature. Note that other works may use different cosmologies and IMFs, with potentially different halo mass definitions and mass accretion models which could lead to biases compared to our results. Hence, 
renormalizing their results to our choice of cosmology and SFR$-$luminosity conversion is non-trivial and we show the SFE curves as presented in their works.

At a glance, one can see that the $1\sigma$ SFE constraints are generally informative up to $M_h\gtrsim10^{10}\msun$ at $z=6$ moving down to $M_h\gtrsim10^9\msun$ at $z=14.5$. At lower halo  masses, the SFE is unconstrained and hence, the degeneracy between $\tilde{f}_{\star,\twoi}$ and $M_0$ allows for a wide range of curves that flatten below the observed mass threshold. The shrinking of SFE priors (grey contours in Figure~\ref{F:SFE_UVLF_constraints}) to posteriors (coloured contours in the same Figure) at the low mass end is due to our implicit assertion that the SFE is continuous across halo masses, which is not a priori guaranteed, but a reasonable assumption \citep[but see also the `bursty' or `high scatter' model of][in their Figure 11]{2024ApJ...961...50S}. Furthermore, the steep contours for SFE and the drop in UVLF at the faint magnitude end (which is redshift dependent) are a physical feature --- this is the direct result of the constraint on high values of minimum circular velocity $V_c$ (equivalently $T_\text{vir}$ or $M_\text{vir}$) for star-formation in DM halos.

At the lowest redshift, $z=6$, we find strong agreement with the SFE magnitude and slope of \citet[][T+18]{2018ApJ...868...92T} and \citet[][SF16]{2016MNRAS.460..417S} to $\lesssim 2\sigma$, both of which predict $f_{\star,\twoi} \propto M_h$ (or equivalently, $M_\star \propto M_h^2$). The former is an empirical redshift-independent model based on informed accretion histories from $N$-body simulations, while the latter uses halo-abundance matching assuming smooth gas accretion. The results are also within $2\sigma$ of the deepest Spitzer and HST-calibrated results from \citet[][S+21]{2021ApJ...922...29S}. Compared to the recent work of \citet[][SL24, in particular their `B21+B22' model]{2024ApJ...961...50S}, although roughly consistent at $3\sigma$, we find a steeper slope than their $f_{\star,\twoi} \propto M_h^{0.56}$ and lower efficiency at the the faint end. The feedback-free burst model from \citet[][L+24, their $\epsilon_\text{max}=0.2$ model]{2023MNRAS.523.3201D,2024A&A...690A.108L} is a factor of few lower than our prediction, which is expected because FFB galaxies are less abundant compared to non-FFB galaxies at this redshift.

At $z=8$, we find an almost exact match with the SFE of \citetalias{2018ApJ...868...92T} and \citetalias{2016MNRAS.460..417S}. We are also $\lesssim 2\sigma$ consistent with \citetalias{2024A&A...690A.108L} across the halo mass range, and with \citetalias{2024ApJ...961...50S} at the high mass end (deviating by a factor of few at the low mass end). The UVLF best-fit model at this redshift is in fact more consistent with latest spectroscopic measurements from JWST \citep{2025ApJ...985...80R}, lying slightly above the B21 data points at the bright end.

At $z=10$, we are consistent with most works at the low mass end, while predicting an intermediate SFE at the high mass end, between the FFB model of \citetalias{2024A&A...690A.108L} and others. At $z>10$, we predict an order of magnitude higher SFEs than \citetalias{2018ApJ...868...92T} and strongly favour the FFB model of \citetalias{2024A&A...690A.108L} quantitatively.

This result recovers the slow evolution of the SFE inferred in pre-JWST HST-calibrated results at $z\lesssim 10$, and enhanced SFE from a rapid evolution at $z \gtrsim 10$ in post-JWST models (a transition consistent with the findings of \citealt{2025arXiv250307590C}, although with a different parameterization). Across the entire redshift range, the SFE slope $\langle\alpha_\star\rangle \approx 1.1$ is a reasonable fit to data (see Table~\ref{tab:SFE_constraints} for exact values and error-bars), which is consistent with \citetalias{2018ApJ...868...92T} within errors. 

In Figure~\ref{F:SFE_z_dependence} we show the evolution of the star-formation efficiency of DM halos at $M_h = 10^{10}\msun$ with redshift, and the $1\sigma$ upper limits for $M_h = 10^9\msun$. We find,
\begin{align*}
    f_{\star,\twoi}(M_h=10^{10}\msun) &=
    \begin{cases}
        1.3^{+0.5}_{-0.8} \% \text{ at } z=6 \text{ (low-$z$ fit)}, \\[.5em]
        1.7^{+0.4}_{-0.9} \% \text{ at } z=9 \text{ (low-$z$ fit)}, \\[.5em]
        8.2^{+2.1}_{-4.0} \% \text{ at } z=12 \text{ (high-$z$ fit)}, \\[.5em]
        20.8^{+3.3}_{-14.5} \% \text{ at } z=15 \text{ (high-$z$ fit)}
    \end{cases}
\end{align*}
while that of $f_{\star,\twoi}(M_h=10^9\msun)$ remains at a few sub-percent level, reaching $\approx 1 \rangeto 2\%$ at the highest redshifts, and is weakly constrained due to the lack of UVLF data at the faint end. Such an increase of $\times10$ in SFE at $M_h=10^{10}\msun$ from the global $1\rangeto3\%$ (at low redshifts) to $10\rangeto30\%$ or higher (at high redshifts) is consistent with observations of compact, super star clusters in nuclear starburst galaxies and giant molecular clouds \citep{1998ApJ...498..541K,2005ApJ...635.1062K,2010ApJ...709..191M,2018ApJ...869..126L,2020ApJ...903...50E,2020MNRAS.491.4573R,2021ApJ...918...76C,2024ApJ...967..133S}\footnote{These high efficiencies are also achieved in numerical simulations of compact giant molecular clouds with high surface densities (\citealt{2018ApJ...859...68K,2021MNRAS.506.5512F}; see also \citealt{2025arXiv250505554W} for interesting discussions on GMC to galaxy scale SFE).}. The increased SFE also quantitatively matches previous results in the literature \citep{2022ApJ...938L..10I,2025arXiv250418618Y} which conclude that high-$z$ galaxy observations are not yet inconsistent with $\Lambda$CDM.

\textit{Thus, the UVLF is a powerful dataset providing the strongest current constraints on early galaxies via the SFE. The data favours a constant SFE at $z\lesssim10$ and a rapidly evolving SFE at $z\gtrsim10$. We conduct the first multi-wavelength analysis utilizing high-$z$ galaxy observations out to $z\approx15$.}

\subsubsection{SFRD evolution}

We show the evolution of the star-formation rate density (SFRD) in Figure~\ref{F:SFRD_constraints}. In order to compare our results directly with the literature, we compute the SFRD by integrating the UVLFs down from a magnitude of $M_\text{UV}=-23\magn$ to $M_\text{UV}=-17\magn$, instead of the simulation volume-averaged SFRD defined in Equation~\ref{eqn:sfrdII}. The latter includes the contribution from faint galaxies hosted in halos above $V_c$, which can be as low as $\sim10^{6}\msun$ at $z=15$ and $\sim10^{5}\msun$ at $z=30$ for low $V_c$ values (Equation~\ref{eqn:mvir}), and contribute significantly to the SFR due to a flattening SFE slope. The luminosities are converted to SFR using the conversion factor $\kappa_\text{UV} = 1.15\times10^{-28}\msun\yr^{-1}\erg^{-1}\second\hz$, consistent with the UVLF modelling described in Section~\ref{uvlf_model}.

By construction in Equation~\ref{eqn:mhigh}, the SFRDs for the low-$z$ and high-$z$ fits are consistent at $z=6$, but diverge at higher redshifts. The low-$z$ fit aligns with literature values based on HST observations at $z \lesssim 10$ \citep[e.g.][]{2015ApJ...803...34B,2015ApJ...810...71F,2016MNRAS.459.3812M}, while the high-$z$ fit is consistent with JWST observations at $z \gtrsim 10 \rangeto 15$ \citep[e.g.][]{2023ApJS..265....5H,2024MNRAS.533.3222D,2025arXiv250100984W}. At $z\gtrsim 15$ however, outside the range of our fits, the tentative observations of \citet{2025arXiv250315594P} suggest even higher SFRDs. The difference between FFB and non-FFB models of \citet[][see their Figure~1 where the curves differ by an order of magnitude]{2024MNRAS.532..149L} is quantiatively similar to the low-$z$ and high-$z$ fits in our analysis. Furthermore, since the SFRD converges at the lowest redshifts — which contribute most to the diffuse X-ray and radio backgrounds — the constraints on the radiative efficiencies $f_X$ and $f_r$, described in the next section, are robust to the choice of the $\beta_\star$ parameter.

\subsection{Constraints on $f_X$ and $f_r$}
\label{results_fX_fr_constraints}

We find that the constraints on the X-ray and radio efficiencies of early galaxies, $f_X$ and $f_r$ as defined in Equation~\ref{fX_equation} and Equation~\ref{fr_equation} respectively, become tighter with the inclusion of UVLF data by breaking their degeneracy with the SFE of galaxies. Figure~\ref{F:TrianglePlot_woUVLF} shows the 2D joint posteriors for $\{\tilde{f}_{\star,\twoi}, f_X, f_r\}$ when UVLF data is included and excluded. In the absence of UVLF data, the parameter $\tilde{f}_{\star,\twoi}$ saturates to $\sim10^{-4}$, allowing for larger values of $f_X$ and $f_r$ \citep[and indeed this matches the inference of][]{2024MNRAS.531.1113P}.

Several different datasets contribute to constraints on the X-ray efficiency of early galaxies $f_X$. CXB data (i.e. Chandra and other satellites) sets upper limits on the diffuse X-ray background disfavouring high values of $f_X$; on the other hand, HERA favours high values as efficient heating would suppress the 21-cm signal \citep[as found previously in][]{2023ApJ...945..124H,2024MNRAS.531.1113P}. This squeezes the prior from either ends of its range resulting in a measurement of $f_X$\footnote{These constraints build upon previous works where only one of the two dataset combinations are used: UVLF + HERA in \citealt{2023ApJ...945..124H}, and UVLF + CXB in \citealt{2025arXiv250203525Z}, both of which give strong constraint on one end of the $f_X$ prior range.}. Furthermore, since the diffuse background scales as $f_X \times \text{SFR}$, the inclusion of UVLF data breaks the degeneracy between $f_{\star,\twoi}$ and $f_X$, anchoring the former to observed galaxies as discussed earlier. The joint analysis then yields a measurement of $f_X$:
\begin{align*}
    \log_{10}\left(f_X\right) &= -0.10^{+1.12}_{-0.31}\left(^{+1.33}_{-2.22}\right) \text{ at 68\% (95\%) CL} \\
    \implies f_X &= 0.8^{+9.7}_{-0.4}\left(^{+16.3}_{-0.8}\right) \text{ at 68\% (95\%) CL}.
\end{align*}
which is strongly consistent with theoretical predictions of low metallicity HMXBs at high-$z$ \citep[$f_X=1$;][]{2013ApJ...776L..31F}. This is indeed a tighter constraint than the one obtained without UVLF data (see Table~\ref{tab:fX_fr_constraints}). Hence, $f_X \gtrsim 11$ is ruled out at $1\sigma$ confidence, $f_X \gtrsim 17$ is ruled out at $2\sigma$ confidence, and $f_X \gtrsim 49$ is completely ruled out ($5\sigma$). At the low $f_X$ end, the constraints are weaker. Although disfavouring low $f_X$ values, we do not rule them out. This is on account of the degeneracy between $\tau$ and $f_X$ (see 2D posterior in Figure~\ref{F:TrianglePlot_Joint_astro}), where an advanced reionization at $z \approx 8$ (i.e. higher $\tau$) suppresses the 21-cm power spectrum enough to allow for weak X-ray heating. Inclusion of reionization-era datasets \citep[e.g.][]{2021MNRAS.506.2390Q} can potentially tighten this constraint further, and shed light on whether weak X-ray efficiencies in early galaxies are indeed allowed.

The upper limits from CRB data (i.e. LWA1 and other telescopes) directly constrain the radio efficiency of early galaxies $f_r$ by disfavouring high values. The upper limits from HERA data provide a secondary constraint disfavouring high values of $f_r$ which would result in bright 21-cm signals\footnote{For both parameters, $f_X$ and $f_r$, SARAS~3 constraints are weak with only a marginal preference for higher $f_X$ and lower $f_r$ values. This is because the contribution of faint high-$z$ galaxies (driving the signal in the SARAS~3 band) to the diffuse X-ray and radio backgrounds is small. Compared to \citet{2024MNRAS.531.1113P}, we probe an order of magnitude lower SFEs, especially for the low mass halos abundant at high-$z$, leading to weaker constraints.}. The joint analysis gives us an upper limit on $f_r$:
\begin{align*}
    \log_{10}\left(f_r\right) &< 1.23\, (2.39) \text{ at 68\% (95\%) CL} \\
    \implies f_r &< 16.9\, (243.7) \text{ at 68\% (95\%) CL}.
\end{align*}
As with the $f_X$ parameter, the UVLF data strengthens the constraints by $\times 10$ at $2\sigma$ (see Table~\ref{tab:fX_fr_constraints}). Thus, in this model, $f_r \gtrsim 17$ is ruled out at $1\sigma$ confidence, $f_r \gtrsim 244$ is ruled out at $2\sigma$ confidence, and $f_r \gtrsim 1800$ is completely ruled out ($5\sigma$). The 2D joint posterior in the $f_X-f_r$ plane, shown in Figure~\ref{F:TrianglePlot_Joint_astro}, has an interesting feature ruling out even lower values of $f_r \gtrsim 100$ at $2\sigma$ confidence for $f_X \lesssim 0.1$. This constraint comes from the HERA data disfavouring large 21-cm signals --- a low heating efficiency (i.e. $f_X$) of galaxies would imply that the matter temperature can cool down for longer, leading to a large 21-cm signal. Hence, $f_r$ cannot be too high as it would further increase the contrast between the matter temperature and radio background temperature, leading to an even larger 21-cm signal. Recently \citet{2024ApJ...970L..25S} derived constraints on $f_r$ from clustering of radio sources (not included here) using the anisotropy upper limits from VLA at $4.9\ghz$ and ATCA at $8.7\ghz$ \citep{2014ApJ...780..112H}, which amount to $f_r(z=7) \lesssim 20$ (although assuming a fixed SFE of $10\%$). Our results here are as stringent as this at $1\sigma$, showing the power of UVLF data. Future works will combine the analysis here with radio source clustering data to further tighten the constraint on $f_r$.~\\

The main caveat of our results here is the assumption of a continuous model for X-ray and radio emission across redshifts. The diffuse X-ray and radio backgrounds are integrated observables, building up over cosmic time, with the largest contribution coming from galaxies at $z\approx6$ where the SFRD constraints are tightest\footnote{Since this is also where the SFRDs converge between the low-$z$ and high-$z$ fit, the $f_X$ and $f_r$ constraints are insensitive to $\beta_\star$.}. Thus, our inference of $f_X$ and $f_r$ parameters is most robust close to $z=6$ and becomes increasingly model-dependent at higher redshifts. Just as there is evidence of SFE varying with $z$ (via $\beta_\star$), the $f_X$ and $f_r$ could indeed be very different for early galaxy populations (e.g. Pop~III galaxies as in \citealt{2025NatAs.tmp..132G} or for metallicity-dependent XRBs as in \citealt{2022MNRAS.513.5097K}). This is a limitation of our analysis, and we leave the exploration of such models to future works.

\textit{By jointly constraining the CXB and CRB data with UVLFs, we have significantly tightened the constraints on the X-ray and radio properties of galaxies in the early Universe. This highlights the importance of a well-calibrated SFE model, which enables flexibility and supports broad priors in poorly probed regimes, such as low-mass and high-redshift galaxies.}

\section{Conclusions}
\label{conclusions}



\begin{table}
    \centering
    \renewcommand*{\arraystretch}{1.5}
    \caption{Marginal 1D constraints on the radiative efficiencies $f_X$ and $f_r$ as seen in Figure~\ref{F:TrianglePlot_Joint_astro} and Figure~\ref{F:TrianglePlot_woUVLF}. The quoted errors are at 68\% (95\%) confidence levels around the weighted mean. We find that the constraints are stronger and more accurate when UVLF data is included.}
    \label{tab:fX_fr_constraints}
    \resizebox{\linewidth}{!}{
    \begin{tabular}{cccc}
    \hline
    \textbf{Parameter} & \textbf{Prior} & \textbf{Joint w/o UVLF} & \textbf{Joint w/ UVLF} \\
    \hline
    $\log_{10}\left(f_X\right)$ & $\mathcal{U}\left(-3,3\right)$ & $0.05^{+1.76}_{-1.07}\left(^{+2.24}_{-2.82}\right)$ & $-0.10^{+1.12}_{-0.31}\left(^{+1.33}_{-2.22}\right)$ \\
    $\implies f_X$ & $\log\mathcal{U}(10^{-3},10^3)$ & $1.1^{+63.1}_{-1.0}\left(^{+192.1}_{-1.1}\right)$ & $0.8^{+9.7}_{-0.4}\left(^{+16.3}_{-0.8}\right)$ \\
    \hline
    $\log_{10}\left(f_r\right)$ & $\mathcal{U}\left(-1,5\right)$ & $\lesssim 1.45\,(3.36)$ & $\lesssim 1.23\,(2.39)$ \\
    $\implies f_r$ & $\log\mathcal{U}(10^{-1},10^5)$ & $\lesssim 28.0\,(2282.6)$ & $\lesssim 16.9\,(243.7)$ \\
    \hline
\end{tabular}
}
\end{table}

In this work, we constrain astrophysical parameters of galaxies in the early Universe combining a suite of 30,000 simulations performed using the code \simcode\ with multi-wavelength observations. The simulations span the redshift range $z=6\rangeto50$, and we vary eight astrophysical parameters: the minimum circular velocity for star-formation in dark matter (DM) halos $V_c$, four star-formation efficiency (SFE) parameters which describe a halo mass and redshift dependent SFE of Pop~II galaxies $\left(\tilde{f}_{\star,\twoi}, M_0, \alpha_\star, \beta_\star\right)$, the ionization efficiency of galaxies $\zeta_\text{ion}$, X-ray emission efficiency of galaxies $f_X$, and the radio emission efficiency of galaxies $f_r$.

In order to constrain these parameters with recent data from HST/JWST, we implement a pixel-level UV luminosity function (UVLF) model as a new observable in \simcode, including an analytic prescription for dust attenutation. We then train neural networks to emulate five key observables as functions of the astrophysical parameters: the 21-cm global signal and power spectrum, the present-day cosmic X-ray background (CXB), the cosmic radio background (CRB), and UVLFs. We perform Bayesian analysis to infer our astrophysical parameters using the following observational data sets: 21-cm global signal non-detection from SARAS~3 \citep{2022NatAs...6..607S}, 21-cm power spectrum upper limits from HERA Phase 1 observations \citep{2023ApJ...945..124H}, CXB measurements from Chandra and other X-ray telescopes \citep{2006ApJ...645...95H}, CRB measurements from LWA1 and other telescopes \citep{2018ApJ...858L...9D}, UVLF measurements from HST at redshifts $z=6\rangeto9$ \citep{2021AJ....162...47B} and JWST at redshifts $z=9\rangeto14.5$ \citep{2024MNRAS.533.3222D}. We also indirectly include the \textit{Planck 2018} measurement of optical depth of CMB photons to reionization $\tau=0.054\pm0.07$ \citep{2020A&A...641A...6P} with a $3\sigma$ prior on it, as a proxy of excluding $\zeta_\text{ion}$ values that lead to very late or very early reionization.

Our joint analysis thus builds on the work of \citet{2024MNRAS.531.1113P} and improves upon it by incorporating the flexible SFE model and inclusion of UVLF data for constraints. Our main findings include:
\begin{enumerate}[1.]

    \item The minimum circular velocity for star-formation in DM halos $V_c$ is constrained by a combination of SARAS~3 data and UVLF data. The UVLF data provides the dominant constraint, disfavouring large $V_c$ values via direction observation of faint magnitude galaxies. This upper limit is weakened by the SARAS~3 data which either disfavours or begins to rule out (depending on the foreground fit) low $V_c$ values due to the absence of a deep 21-cm absorption signal in their observed redshift band $z=1 5 \rangeto 25$. The joint analysis with and without the SARAS~3 data yields $V_c \lesssim 21.3\km\second^{-1}$ and $V_c \lesssim 25.2 \km\second^{-1}$ at $68\%$ confidence respectively. We thus demonstrate the synergy that exists between 21-cm signal (even non-detections) and high-$z$ galaxy observations in tightening the constraints on this parameter, improving upon works that only use either UVLF data \citep{2024ApJ...961...50S,2025arXiv250203525Z} or 21-cm signal data \citep{2024MNRAS.531.1113P}. The main challenge in the analysis is the degeneracy between $V_c$ and the log-log polynomial foreground model used to fit SARAS~3 data. Using physically motivated foregrounds, future works can tighten this constraint further.
    
    \item The star-formation efficiency model, as illustrated in Figure~\ref{F:SMHM}, consists of four parameters: $\tilde{f}_{\star,\twoi}$ scales the magnitude of the SFE, $M_0$ sets the halo mass below which the curve flattens, $\alpha_\star$ is the power law slope of the $M_h$-dependence, and $\beta_\star$ controls the redshift-dependence anchored to $z=6$. These parameters are almost entirely constrained by the UVLF data with other data sets providing much weaker constraints. The data favours a non-monotonic evolution of the SFE with redshift: slow evolution at $z\approx 6\rangeto10$ with $\beta_\star \lesssim 1.4$ (and the posterior saturating to zero, consistent with no evolution), and then a rapid evolution at $z\approx10-15$ with $\beta_\star = 3.6^{+0.7}_{-0.7}$ at 68\% confidence. The joint fit combining both redshift bins gives a maximum posterior estimate $\beta_\star = 2.9^{+0.9}_{-0.7}$. Thus, we find consistency with pre-JWST HST-calibrated results at low redshifts \citep{2016MNRAS.460..417S,2018ApJ...868...92T}, and post-JWST enhanced SFE models at high redshifts (in particular the feedback-free burst model of \citealt{{2023MNRAS.523.3201D,2024A&A...690A.108L}}, and the density-modulated SFE of \citealt{2025arXiv250505442S}). The slope of the SFE with halo mass $\alpha_\star = 1.1^{+0.2}_{-0.2}$ is consistent across the two redshift bins for $M_h\gtrsim10^{10}\msun$, in agreement  with the findings of \citet{2018ApJ...868...92T}. Futhermore, under our composite model, we infer that DM halos of mass $M_h = 10^{10}\msun$ have a SFE of $1\rangeto2\%$ at $z\lesssim10$, $8\%$ at $z=12$ and $21\%$ at $z=15$ (for the full functional posteriors of the SFE in the low and high-redshift regime, see Figure~\ref{F:SFE_UVLF_constraints}).
    
    \item The X-ray efficiency of early galaxies $f_X$, defined relative to the theoretical predictions of low metallicity high mass X-ray binaries \citep[HMXBs;][]{2013ApJ...776L..31F} in Equation~\ref{fX_equation}, is predominantly constrained by a combination of CXB and HERA data. CXB data disfavour very efficient sources by their direct contribution to the diffuse background seen today. HERA disfavours very inefficient X-ray sources, as late heating would lead to stronger fluctuations in the 21-cm brightness temperature at $z=7\rangeto11$. The joint analysis gives us a measurement of $f_X = 0.8^{+9.7}_{-0.4}\left(^{16.3}_{-0.8}\right)$ at $68\%$ ($95\%$) confidence levels. We rule out $f_X \gtrsim 17$ at $2\sigma$ confidence and $f_X \gtrsim 49$ at $5\sigma$ confidence level. Our findings imply that the X-ray properties of early Pop~II galaxies at $z\gtrsim6$ 
    are consistent with predictions of low-metallicity HMXBs from \citet{2013ApJ...764...41F}. Note that this constraint becomes increasingly model-dependent at higher redshifts, since the CXB is an integrated observable (with largest contribution from galaxies at $z\sim6$) and we assume a constant model for X-ray emission across redshifts \citep[see, e.g.][for an alternate redshift-evolving model]{2022MNRAS.513.5097K}.
    
    \item The radio efficiency of early galaxies $f_r$, defined relative to those in the local Universe in Equation~\ref{fr_equation}, is constrained by a combination of CRB and HERA data (with a weak contribution from SARAS~3). CRB data disfavours high radio efficiency of galaxies by their direct contribution to the diffuse background seen today. HERA further disfavours strong radio sources since they would boost the amplitude of 21-cm signal fluctuations beyond observed upper limits. The joint analysis results in an upper limit $f_r \lesssim 16.9\,(243.7)$ at $68\%$ ($95\%$) confidence levels. We rule out $f_r \gtrsim 1800$ at $5\sigma$ confidence level. At $1\sigma$, this is comparable to the constraints obtained by \citet{2024ApJ...970L..25S} using radio source clustering data from VLA and ATCA (not included here).
    
    \item The joint analysis without UVLF data provides weaker constraints on $f_X$ and $f_r$ due their degeneracy with an unconstrained SFE in the calculation of the diffuse backgrounds (e.g. disfavouring $f_X\gtrsim193$ and $f_r\gtrsim2280$ at $2\sigma$). By including UVLF data and thus informing the model of SFE in observed galaxies, we break this degeneracy and tighten the constraints by a factor of $\sim 10$ \citep[disfavouring $f_X\gtrsim 17$ and $f_r\gtrsim 244$ at $2\sigma$; see also][whose result supports this conclusion]{2025arXiv250203525Z}. 
\end{enumerate}

In conclusion, we have shown the critical role of UVLFs in our understanding of the Cosmic Dawn and Epoch of Reionization: they set an upper limit on the minimum mass of star-forming halos, constrain the star-formation efficiency of galaxies across cosmic epochs, and thus also break degeneracies with other astrophysical parameters constrained by different wavelength probes.

Exciting prospects remain for future works on multi-wavelength inferences. The analysis done here can be extended to include other reionization era data sets such as Lyman-line, Ly$\alpha$ forest and quasar observations to constrain ionizing properties of galaxies \citep[][]{2021MNRAS.506.2390Q,2025PASA...42...49Q,2025arXiv250409725S}, flexible Pop~III models to constrain their role during the Cosmic Dawn \citep[][]{2024MNRAS.529..519G,2024MNRAS.531.1113P,2025arXiv250203525Z}, as well as deeper JWST data \citep[e.g. $z=15-25$,][published during the final stages of preparation of this manuscript]{2025arXiv250315594P}. Furthermore, future synergies using cross-correlations with CMB \citep{2022ApJ...928..162L}, line intensity mapping experiments \citep[see, e.g.][for a recent review]{2022A&ARv..30....5B} and the cosmic near-infrared background \citep[CNIRB;][]{2014ApJ...790..148M,2021MNRAS.508.1954S} will enable a comprehensive multi-wavelength multi-scale picture of the early Universe IGM and ISM  in the coming years.

\section*{Acknowledgements}

JD would like to thank Adam Ormondroyd for his help in understanding and using \polychord, Will McClymont for discussions on the astrophysics of high-$z$ galaxies, and Clàudia Janó Muñoz for comments on the first draft. JD would also like to thank the anonymous reviewer for their constructive feedback which has improved the manuscript.

JD acknowledges support from the Boustany Foundation and Cambridge Commonwealth Trust in the form of an Isaac Newton Studentship. TGJ and EdLA acknowledge the support of the Science and Technology Facilities Council (STFC) through grant numbers ST/V506606/1 and a Rutherford Fellowship, respectively. HTJB would like to acknowledge support from the Kavli Institute for Cosmology Cambridge, and the Kavli Foundation. SP acknowledges the support of the Cambridge Commonwealth, European \& International Trust and the Cavendish Laboratory through a Cambridge International \& Isaac Newton Studentship. ST acknowledges support by the Royal Society Research Grant G125142. RB acknowledges the support of the Israel Science Foundation (grant No. 1078/24). The simulations and analysis were performed under DiRAC project number APP15468 using the DiRAC@Durham facility managed by the Institute for Computational Cosmology on behalf of the STFC DiRAC HPC Facility (www.dirac.ac.uk). The equipment was funded by BEIS capital funding via STFC capital grants ST/P002293/1, ST/R002371/1 and ST/S002502/1, Durham University and STFC operations grant ST/R000832/1. DiRAC is part of the National e-Infrastructure.

\section*{Data Availability}

The 30,000 simulations performed using \simcode, the trained emulators, the \polychord\ chains, and other analysis scripts used in this work can all be made available upon reasonable request to the corresponding author.


\bibliographystyle{mnras}
\bibliography{SFE_constraints}


\appendix

\section{Impact of fixed UVLF parameters}
\label{uvlf_variation}

\begin{figure}
	\includegraphics[width=\linewidth]{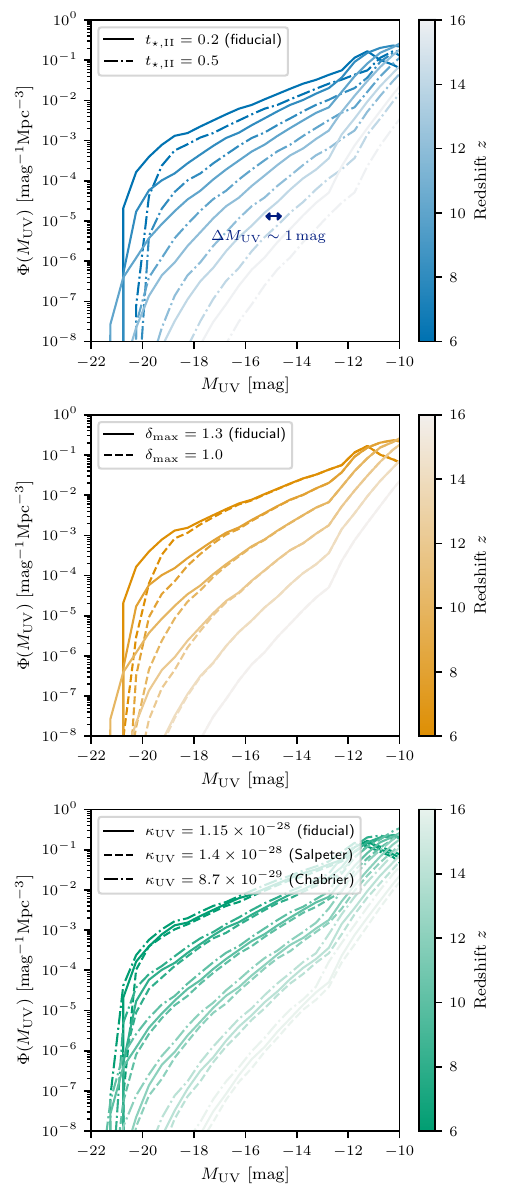}
	\caption{Assessing the impact of the fixed parameters in our UVLF modelling. \textbf{Top}: Changing the averaging time-scale $t_{\star,\twoi}$ from $0.2H(z)^{-1}$ (fiducial) to $0.5H(z)^{-1}$ shifts the UVLF by $\Delta M_\text{UV} = 1\magn$. This parameter is completely degenerate with $\tilde{f}_{\star,\twoi}$. \textbf{Middle}: Varying the threshold for linear overdensities $\delta_\text{max}$ in the simulation affects the abundance of bright galaxies. The effect is most pronounced at low redshifts. \textbf{Bottom}: The effect of varying the SFR-$L_\text{UV}$ conversion factor $\kappa_\text{UV}$, from the fiducial value based on \citet{2014ARA&A..52..415M}, to a fixed-metallicity Salpeter and Chabrier IMF. The effect is sub-dominant to the impact of $t_{\star,\twoi}$ or $\delta_\text{max}$.}
	 \label{F:UVLF_variation}
\end{figure}

\let\oldthefigure\thefigure
\setcounter{figure}{0}
\renewcommand{\thefigure}{B\arabic{figure}}
\begin{figure*}
	\includegraphics[width=\linewidth]{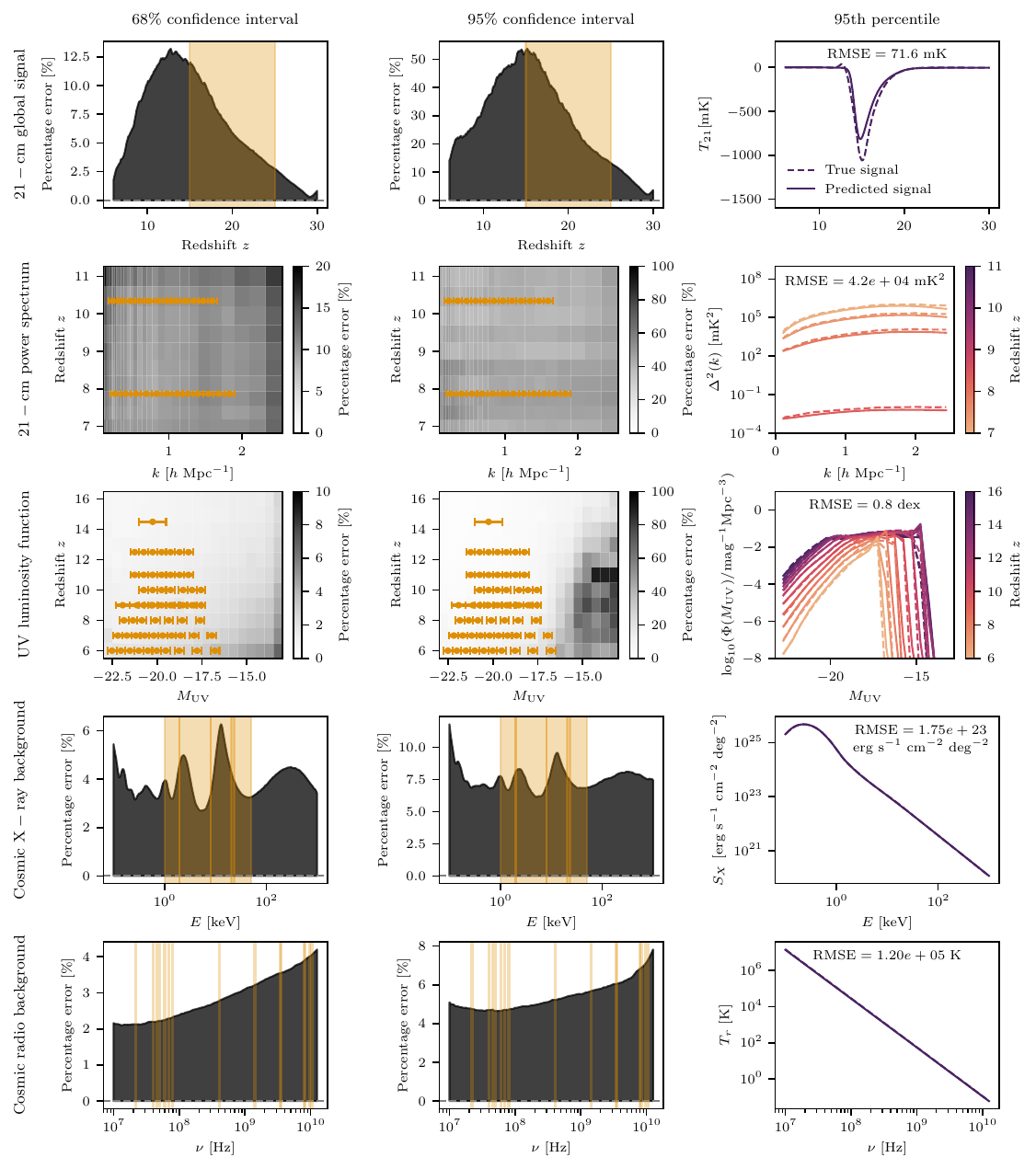}
	\caption{Errors on the 1D emulators for the 21-cm global signal, cosmic X-ray background, cosmic radio background, and 2D emulators for the 21-cm power spectrum and UV luminosity functions, as a function of the input space (e.g., $z$ for the global signal) shown at the 68\% and 95\% confidence levels calculated using the test data. The third column shows the test data alongside its emulated counterpart, at the 95th percentile of RMSE. The data and shaded regions in yellow show the observational data used in the constraints, for comparison.}
	 \label{F:Emulator_errors}
\end{figure*}
\let\thefigure\oldthefigure

\let\oldthetable\thetable
\setcounter{table}{0}
\renewcommand{\thetable}{C\arabic{table}}
\begin{table*}
    \renewcommand\arraystretch{1.1}
     \caption{Summary of works on astrophysical constraints using the code \simcode. The astrophysical parameters used in each work are listed, along with the observational data used. For details on the parameters, please refer to the cited works --- \textit{the model changes between works, but the qualitative interpretation of parameters is similar}. Previous works use a constant Pop~II SFE in the atomic cooling regime ($f_\star$ or $f_{\star,\twoi}$). In this work, we modify and introduce halo mass and redshift-dependent SFE with the efficiency level normalized to $\tilde{f}_{\star,\twoi}$. We also introduce new parameters $\alpha_\star$, $M_0$ and $\beta_\star$ (see Equation~\ref{eqn:sfe_model}). The $^\dagger$ indicates improved upper limits from HERA Phase 1 \citep{2023ApJ...945..124H}, as opposed to the earlier limits \citep{2022ApJ...925..221A}.}
     \label{tab:other_works}
     \centering
     \resizebox{\textwidth}{!}{
     \begin{tabular}{cccc}
        \hline
      \textbf{Work} & \textbf{Focus of the work} & \textbf{Astrophysical parameters} & \textbf{Observational data} \\
        \hline 
        \citet{2017MNRAS.464.3498F} & CXB constraint & $V_c, \tau, \alpha_X, E_\text{min,X}, f_X$ & CXB \\
        \citet{2017ApJ...845L..12S} & SARAS~2 first analysis & $V_c, f_{\star}, \tau, R_\text{mfp}, \alpha_X, E_\text{min,X}, f_X$ & SARAS~2 \\
        \citet{2018ApJ...858...54S} & SARAS~2 reanalysis & $V_c, f_{\star}, \tau, R_\text{mfp}, \alpha_X, E_\text{min,X}, f_X$ & SARAS~2 \\
        \citet{2019ApJ...875...67M} & EDGES High-Band analysis & $V_c, f_{\star}, \tau, R_\text{mfp}, \alpha_X, E_\text{min,X}, f_X$ & EDGES \\
        \citet{2020MNRAS.498.4178M} & LOFAR excess radio constraints & $V_c, f_{\star}, \tau, R_\text{mfp}, A_r$ & LOFAR \\
        \citet{2022MNRAS.513.4507B} & SARAS 2 Bayesian analysis & $V_c, f_{\star}, \tau, R_\text{mfp}, \alpha_X, E_\text{min,X}, f_X, f_r$
        & SARAS 2 \\
        \citet{2022ApJ...924...51A} & HERA excess radio constraints
        & $V_c, f_\star, \tau, f_X, f_r, A_r$ & HERA P1 \\
        \citet{2022NatAs...6.1473B} & SARAS 3 Bayesian analysis & $V_c, f_{\star}, \tau, R_\text{mfp}, \alpha_X, E_\text{min,X}, f_X, f_r, A_r$ & SARAS 3 \\
        \citet{2023ApJ...945..124H} & HERA excess radio constraints & $V_c, f_\star, \tau, R_\text{mfp}, \alpha_X, \nu_\text{min,X}, f_X, f_r$ & HERA P1$^\dagger$\\
        \citet{2024MNRAS.527..813B} & 21-cm joint constraint & $V_c, f_\star, \tau, f_X, f_r$ & HERA P1, SARAS~2/3, LOFAR, MWA \\
        \citet{2024MNRAS.529..519G}
        &  Cosmic string constraints & $V_c, f_{\star,\twoi}, f_{\star,\threei}, t_\text{delay}, \tau, \alpha_X, E_\text{min,X}, f_X, A_r$ & HERA P1$^\dagger$, SARAS~3, CXB  \\
        \citet{2024MNRAS.531.1113P} & Multi-wavelength synergies & $V_c, f_{\star,\twoi}, f_{\star,\threei}, t_\text{delay}, \tau, \alpha_X, E_\text{min,X}, f_X, f_r$ &  HERA P1$^\dagger$, SARAS~3, CXB, CRB \\
        \citet{2024ApJ...970L..25S} & Radio galaxy clustering constraints & $f_r, f_X$ & CRB, ATCA, VLA \\ 
        \textbf{This work} & \textbf{Enhanced SFE and ULVFs} & $\mathbf{V_c, \tilde{f}_{\star,\twoi}, \tau, M_0, \alpha_\star, \beta_\star, f_X, f_r}$ & \textbf{HERA P1$^\dagger$, SARAS 3, CXB, CRB, UVLF} \\
        \hline
     \end{tabular}
    }
\end{table*}
\let\thetable\oldthetable

Here, we briefly assess the impact of the fixed but uncertain parameters such as $t_{\star,\twoi}$, $\delta_\text{max}$ and $\kappa_\text{UV}$ on the UV luminosity function.

\begin{itemize}
    \item \textbf{Impact of $t_{\star,\twoi}$}: The effect of the averaging time-scale $t_{\star,\twoi}$ on the UVLF is to shift the entire distribution by a fixed $\Delta M_\text{UV} = 2.5\times \Delta\log(t_{\star,\twoi})$. This is because the UVLF is a direct tracer of the SFR, which is scaled by $t_{\star,\twoi}$. In Figure~\ref{F:UVLF_variation}, we see the effect of varying $t_{\star,\twoi} = 0.2\rightarrow0.5$ resulting in a shift in magnitude $\Delta M_\text{UV} = 1\magn$. The parameter is completely degenerate with $\tilde{f}_{\star,\twoi}$, introduced in Equation~\ref{eqn:sfe_model}, and can be subsumed into its interpretation (see Footnote~\ref{tstar_footnote}). Since we typically account for an error magnitude of 20\% from semi-numerical modelling in our Bayesian analysis, the effect of $t_{\star,\twoi}$ (varied by a factor of $2-3$) is within the error budget of our inference.

    \item \textbf{Impact of $\delta_\text{max}$}: Next, we find that the UVLF is sensitive to the overdensity cap $\delta_\text{max}$, which determines the largest overdensity allowed in the simulation, in a notable way. It directly affects the abundance of the brightest galaxies in the simulation, and we attempt to set this parameter in an informed manner for our adopted hybrid HMF (as described in Footnote~\ref{deltamax_footnote}) to $\delta_\text{max}=1.3$. The effect is apparent at low redshifts and becomes less pronounced at $z\gtrsim10$ as overdensities are less evolved. We leave the development of structure-formation in the non-linear regime within the framework of our semi-numerical simulations to a future work.

    \item \textbf{Impact of $\kappa_\text{UV}$}: Finally, the effect of assuming a fixed metallicity Salpeter IMF or a Chabrier IMF to calculate the SFR-$L_\text{UV}$ conversion factor $\kappa_\text{UV}$, compared to the \cite{2014ARA&A..52..415M} value, proves to be sub-dominant compared to the effect of $\delta_\text{max}$ or $t_{\star,\twoi}$. Note that we do not compare to Pop~III top-heavy IMF, which can indeed have an elevated $\kappa_\text{UV}$, instead choosing to focus our inference on Pop-II type star-formation which is generally dominant at the redshifts of interest \citep[$z<15$, e.g.][]{2018MNRAS.479.4544M,2019MNRAS.488.2202J,2020ApJ...897...95V,2022ApJ...936...45H}. Furthermore, \citet{2024A&A...686A.138C} suggested that a top-heavy IMF alone is insufficient to explain the observed UVLFs on account of the resulting strong stellar feedback.
    
\end{itemize}

\section{Emulator accuracy}
\label{appendix_emuaccuracy}

Here, we analyze the accuracy of our emulators in predicting the observables of interest, and describe how we incorporate this into our likelihoods. Often, the test data is used as a validation set to calculate the loss of an emulator (whether it is root-mean squared error, or mean absolute error). For a 1D observable target function $f(x|\theta)$ such as the 21-cm global signal $T_{21}(z|\theta_\text{ast})$, where $x$ is the `input space' and $\theta$ is the parameter space, these metrics are calculated by averaging all test data points, and across the input space. However, we can extend this idea to characterise the error as a function of the input space $x$, instead of a fixed average value:
\begin{equation}
    \epsilon(x|\theta_i) = \left|\dfrac{f_\text{theory}(x|\theta_i) - f_\text{emu}(x|\theta_i)}{f_\text{theory}(x|\theta_i)}\right|.
    \label{eqn:emulator_error}
\end{equation}
where $\theta_i$ is the $i$-th parameter set in the test data, $f_\text{theory/emu}$ are the theory (simulated) and emulated observables respectively. Instead of averaging, we can sort the errors (over $\theta_i$, keeping $x$ bins separate) in ascending order and define 68\% and 95\% confidence intervals which yields $\epsilon_{1\sigma}(x)$ and $\epsilon_{2\sigma}(x)$ respectively. This gives us an idea of where in the $x$-space the emulator is performing well, and provides a way to accurately account for this error in final analysis. This method has been previously explored in the context of posterior validation for 21-cm emulators \citep{2023ApJ...959...49D,2024MNRAS.527.9833B} and for CMB emulators \citep{2024MNRAS.531.1351B}. In particular, its utility in our analysis is to check whether the emulator is performing well in the region of the input space we care about the most --- i.e., where the observational data lies. For example, in case of the 21-cm signal, we are interested in the SARAS~3 band of $z=15\rangeto25$. The emulator accuracy outside this range is less important. This can be easily extended to 2D functions like 21-cm power spectrum $\Delta_{21}^2(k,z|\theta_\text{ast})$ or UVLFs $\Phi(M_\text{UV},z|\theta_\text{ast})$.

Figure~\ref{F:Emulator_errors} shows the errors on all emulators at the 68\% and 95\% confidence levels, alongside the observational data. Note that, since the global signal potentially crosses zero, we define the error metric in Equation~\ref{eqn:emulator_error} with $\text{max}(T_{21,\text{true}}(z|\theta_\text{ast}))$ instead of $T_{21,\text{true}}(z|\theta_\text{ast})$ in the denominator. In principle, this metric can be defined in a variety of ways (using squared error instead of absolute, for example).

Once the emulator error is characterised in this way, we calculate the total `model error' as the incoherent sum of both simulation error and emulator error:
\begin{equation}
    \sigma_\text{model}(x) = f(x) \times \sqrt{\epsilon_\text{theory}^{2} + \epsilon_{\text{emu},1\sigma}^{2}(x)} \\
\end{equation}
where $\epsilon_\text{theory}=20\%$ is the assumed error on the semi-numerical model implemented in \simcode. This is then used to calculate the likelihood of the data given the model, as described in Section~\ref{ss:likelihoods}.

\section{Extra tables and figures}
\label{extra_figures}

Table~\ref{tab:other_works} shows a summary of past works constraining astrophysics using \simcode\, listed for ease of parsing literature and comparing results.

\begin{figure*}
	\includegraphics[width=\linewidth]{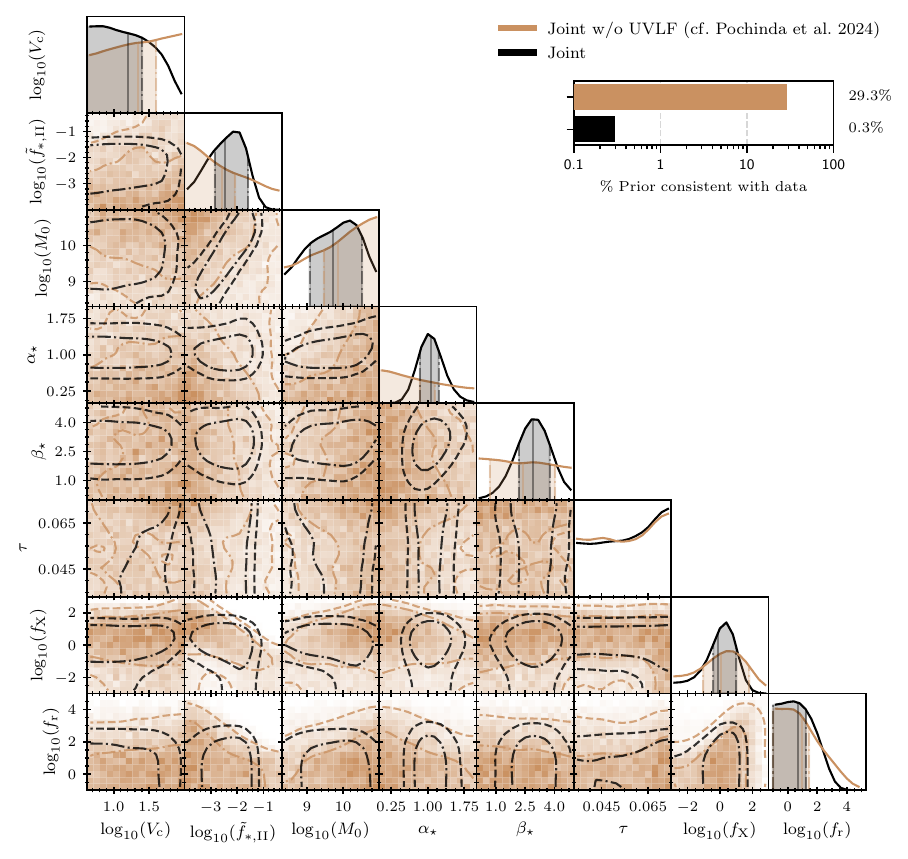}
	\caption{Similar to Figure~\ref{F:TrianglePlot_Joint_astro}, but showing joint constraints with and without the UVLF dataset. The 2D posterior contours show the 68\% (dash-dotted) and 95\% credible regions (dashed), with filled contours for the case without UVLF data. The 1D marginal posterior PDFs show the weighted mean (solid line) and 68\% credible region (shaded with dash-dotted outline). The top right panel shows the prior to posterior volume contraction for each dataset. Note in particular, without the UVLF dataset: $V_c$ saturates to upper prior limit, the four SFE parameters are consistent with a constant, fixed value at the lower limit of $0.01\%$ \citep[recovering the model used in][and indeed their results]{2024MNRAS.531.1113P}, and $f_X$ and $f_r$ are weakened by their degeneracy with SFE.}
	 \label{F:TrianglePlot_woUVLF}
\end{figure*}

Figure~\ref{F:TrianglePlot_woUVLF} shows the joint constraints on the astrophysical parameter space with and without the UVLF dataset, for comparison with the joint analysis in Figure~\ref{F:TrianglePlot_Joint_astro}. Although the constraints are model dependent, the case without UVLF data is consistent with the inference of \citet{2024MNRAS.531.1113P} who assume a constant SFE in the atomic cooling regime. This is a special case of our more general model, where the SFE depends on both halo-mass and redshift.

\begin{figure*}
	\includegraphics[width=\linewidth]{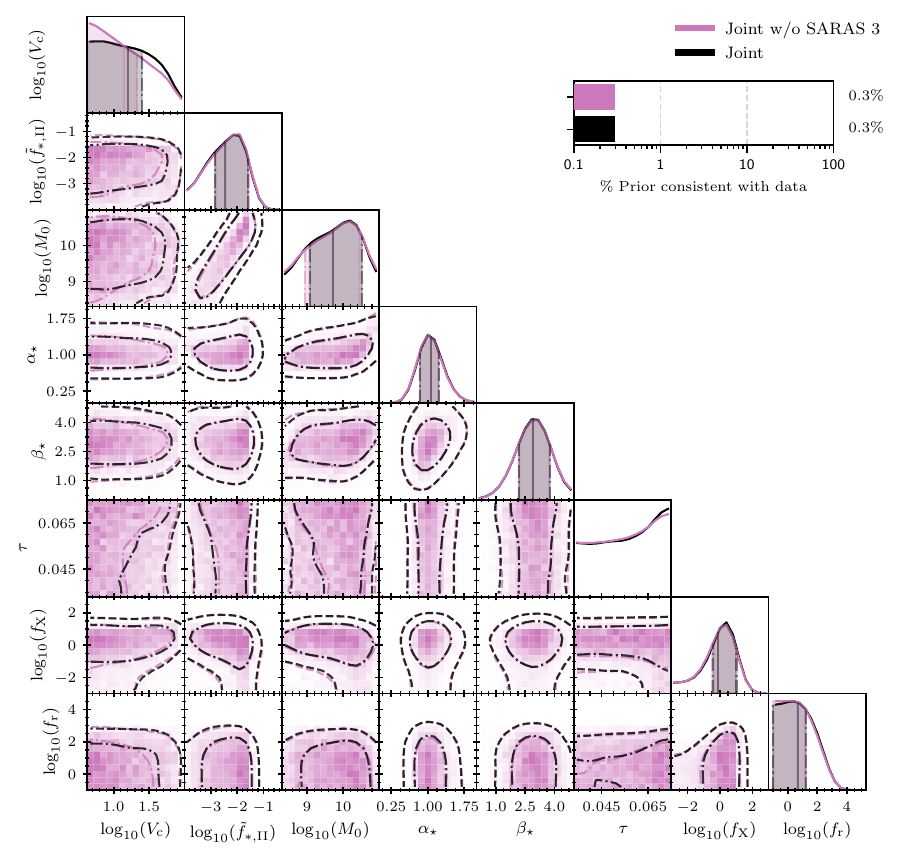}
	\caption{Similar to Figure~\ref{F:TrianglePlot_Joint_astro}, but showing joint constraints with and without the SARAS~3 dataset. The 2D posterior contours show the 68\% (dash-dotted) and 95\% credible regions (dashed), with filled contours for the case without SARAS~3 data. The 1D marginal posterior PDFs show the weighted mean (solid line) and 68\% credible region (shaded with dash-dotted outline). The top right panel shows the prior to posterior volume contraction for each dataset. The SARAS~3 data weakly constrains the minimum circular velocity $V_c$ at the lower end of the prior (as opposed to the UVLF data which disfavours the higher end). The constraint is highly sensitive to the SARAS~3 foreground fit as discussed in Section~\ref{results_Vc_constraints} (see, e.g. Figure~\ref{F:TrianglePlot_binsZ} where the constraint is strongest, and the synergy between UVLFs and SARAS~3 is most visible).}
	 \label{F:TrianglePlot_woSARAS}
\end{figure*}

Figure~\ref{F:TrianglePlot_woSARAS} shows the joint constraints on the astrophysical parameter space with and without the SARAS~3 dataset, for comparison with the joint analysis in Figure~\ref{F:TrianglePlot_Joint_astro}. We demonstrate the synergy between SARAS~3 (or more generally, 21-cm global signal limits) and UVLF data in constraining the minimum circular velocity $V_c$ for star-formation in DM halos.

\begin{figure*}
    \includegraphics[width=\textwidth]{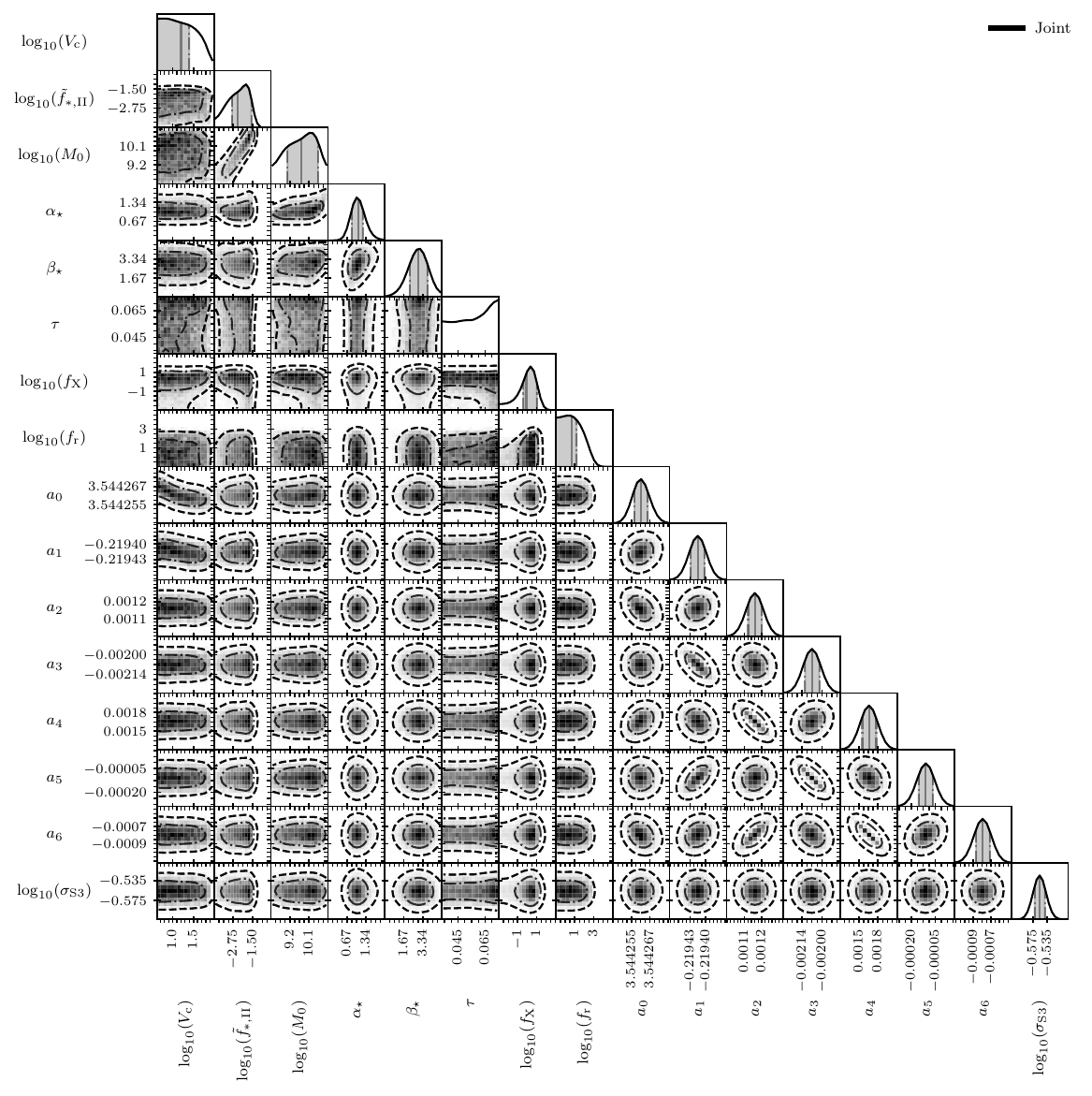}
    \caption{Similar to Figure~\ref{F:TrianglePlot_Joint_astro}, but showing the constraints on the full parameter space (i.e. astrophysical parameters + SARAS~3 nuisance parameters) for the joint analysis. The 2D contours show the 68\% (dash-dotted) and 95\% (dashed) credible regions, while the 1D marginal posterior PDFs show the weighted mean (solid line) and 68\% credible region (shaded with dash-dotted outline). Note in particular, the constraint on the miniminum circular velocity $V_c$ has some degeneracy with the lowest order coefficients of the SARAS~3 foreground polynomial, $a_0$ and $a_1$ (which have the largest effect).}

    \label{F:TrianglePlot_Joint_full}
\end{figure*}

Figure~\ref{F:TrianglePlot_Joint_full} shows the constraints on the full parameter space (i.e. astrophysical parameters + SARAS~3 nuisance parameters) for the joint analysis. We draw attention particularly to the degeneracy between the minimum circular velocity $V_c$ and the lowest order polynomial coefficients of the SARAS~3 foreground polynomial, $a_0$ and $a_1$, which is explained in more detail in Section~\ref{results_Vc_constraints}.

    \bsp	
\label{lastpage}
\end{document}